\documentclass[aps,twocolumn,floatfix,titlepage,balancelastpage,raggedbottom,amsfonts,amssymb,amsmath,showkeys,superscriptaddress]{revtex4}  
\usepackage{epsfig}
\usepackage{bm}
\usepackage{times}
\usepackage{mathptmx}
\usepackage{amsmath}
\usepackage{graphicx}
\usepackage{graphics}
\usepackage{amssymb}
\usepackage{bm}
\usepackage[english]{babel}
\begin{document}
\title{Surface growth for molten silicon infiltration into carbon millimeter-sized channels: Lattice-Boltzmann simulations, experiments and models}
\author{Danilo Sergi}
\affiliation{University of Applied Sciences SUPSI, 
The iCIMSI Research Institute, 
Galleria 2, CH-6928 Manno, Switzerland}
\author{Antonio Camarano}
\affiliation{University of Alicante,
Department of Inorganic Chemistry, 
Apdo.~99, 03080 Alicante, Spain}
\author{Jos\a'e Miguel Molina}
\affiliation{University of Alicante,
Department of Inorganic Chemistry, 
Apdo.~99, 03080 Alicante, Spain}
\author{Alberto Ortona}
\affiliation{University of Applied Sciences SUPSI, 
The iCIMSI Research Institute, 
Galleria 2, CH-6928 Manno, Switzerland}
\author{Javier Narciso}
\affiliation{University of Alicante,
Department of Inorganic Chemistry, 
Apdo.~99, 03080 Alicante, Spain}

\date{\today}


\begin{abstract}
The process of liquid silicon infiltration is investigated for channels with radii from $0.25$ to $0.75$ [mm] drilled in compact 
carbon preforms. The advantage of this setup is that the study of the phenomenon results to be simplified. For comparison purposes, 
attempts are made in order to work out a framework for evaluating the accuracy of simulations. The approach relies on dimensionless numbers
involving the properties of the surface reaction. It turns out that complex hydrodynamic behavior derived from second Newton law can be made
consistent with Lattice-Boltzmann simulations. The experiments give clear evidence that the growth of silicon carbide proceeds in two
different stages and basic mechanisms are highlighted. Lattice-Boltzmann simulations prove to be an effective tool for the description
of the growing phase. Namely, essential experimental constraints can be implemented. As a result, the existing models are useful to gain
more insight on the process of reactive infiltration into porous media in the first stage of penetration, i.e.~up to pore closure because
of surface growth. A way allowing to implement the resistance from chemical reaction in Darcy law is also proposed.
\end{abstract}
\maketitle

\section{Introduction}

Reactive molten silicon (Si) infiltration into carbon (C) preforms is a process widely employed for the manufacturing of
ceramic components designed for advanced applications \cite{hillig,krenkel,brake,gadow,salamone,paik}. The liquid is introduced into
the porous structure by capillary forces. This is possible because Si reacts with C to form silicon carbide (SiC) 
\cite{mortensen,dezellus,dezellus2,highT,passerone,nisi,nisi2,htc09,einset1,einset2}. 
Indeed, Si wets SiC but it does not wet C. Another important property of this reaction is that the SiC formation leads to surface growth. 
The thickening of the surface behind the invading front results in a retardation of the infiltration process and it can also stop the 
fluid motion when pore closure occurs. On the other hand, the SiC phase is a desired property since providing hardness to the material.
Furthermore, the surface reaction influences the hydrodynamic behavior of the melt significantly, since the infiltration dynamics 
does not follow the usual Washburn behavior but it displays a linear dependence on time \cite{nisi,nisi2,htc09}.

In this study, the experiments of infiltration are performed with channels of millimeter width made in dense carbon preforms.
This is done with the purpose to gain insight into the process of surface growth inside porous media at the micron scale.
The advantage of the ideal system of a single channel is to retain the basic mechanisms underlying the surface reaction without
the influence of the porous structure. The experiments clearly show that SiC formation is determined by diffusion first
through the liquid and then through the solid SiC layer, much slower. Of course, the complexity of the phenomenon  goes beyond
the capabilities of the models for the simulations used here (see Sec.~\ref{sec:model}). In order to elucidate this aspect,
the role of the prior C dissolution and pockets formation is discussed in detail. Nevertheless, it turns out that the models for
the simulations proposed in this work can treat the process of surface growth leading to pore closure. This is important
for the optimization of Si-infiltrated ceramic composites and their manufacturing.

The process of infiltration is simulated using the Lattice-Boltzmann (LB) method. This approach is based on the
discretization of the velocity space, besides the usual discretization of the physical space. The hydrodynamic behavior can be
reproduced in the incompressible limit at small Mach numbers. An increasing variety of problems can be worked out exhaustively
by this method \cite{book1,benzi,book2,wolfram,review}. To the end of comparison with experiments, we propose a framework
for assessing the accuracy of simulations. The method is based on dimensionless numbers taking into account the effects
of the surface reaction.  
Interestingly, some properties of experimental systems can be addressed after a suitable calibration. Notably, real data for surface
growth can be imposed. The outcome of simulations proves to be in line with experimental observations for SiC formation. In
general, this work allows to make more precise the predictive power of LB simulations for reactive infiltration thanks to ad
hoc experiments. These results are valuable in order to sharpen the design and interpretation of further research aiming at a
direct comparison with experimental work.


\section{Mathematical background}

\begin{figure}[t]
\includegraphics[width=8.4cm]{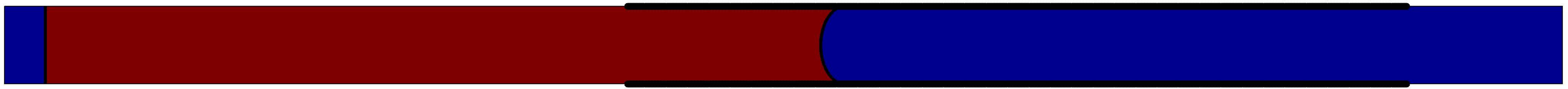}
\caption{\label{fig:channel}
Channel infiltrated by a liquid. The liquid is represented in red; in blue the vapor phase. The solid phase is shown in
dark. For LB simulations, the boundaries of the simulation domain are periodic, reproducing a flat interface characteristic
of a large reservoir.}
\end{figure}

Figure \ref{fig:channel} shows capillary infiltration into a channel. In 3D this phenomenon is described by
the differential equation \cite{chibarro,pianese}
\begin{multline}
\pi r^{2}[\rho_{\mathrm{g}}(L-z)+\rho_{\mathrm{l}}z]\ddot{z}+\pi r^{2}(\rho_{\mathrm{l}}-\rho_{\mathrm{g}})(\dot{z})^{2}=
\\2\pi r\gamma\cos\theta-8\pi[\mu_{\mathrm{g}}(L-z)+\mu_{\mathrm{l}}z]\dot{z}\ .
\label{eq:capillary}
\end{multline}
The terms on the left hand side come from the inertial forces. The first term on the right hand side
is due to capillary forces and the second one accounts for viscous forces.
$z$ is the centerline position of the invading front. $r$ and $L$ are the radius and the length of the channel.
$\rho$ and $\mu$ indicate the density and the dynamic viscosity; the subscripts l and g stand for liquid and gas.
$\gamma$ is the surface tension and $\theta$ the contact angle. The behavior described by the well-known Washburn equation 
is obtained in the limit $\rho_{\mathrm{g}}/\rho_{\mathrm{l}}\rightarrow 0$ and  $\mu_{\mathrm{g}}/\mu_{\mathrm{l}}\rightarrow 0$. 
In so doing Eq.~\ref{eq:capillary} becomes
\begin{equation}
\ddot{z}=-\frac{(\dot{z})^{2}}{z}+\frac{2\gamma\cos\theta}{\rho r}\frac{1}{z}-\frac{8\mu}{\rho r^{2}}\dot{z}\ .
\label{eq:washburn}
\end{equation}
A solution to this differential equation is given by \cite{washburn}
\begin{equation}
z=\sqrt{ \frac{\gamma r\cos\theta}{2\mu}t+z_{0}^{2} }\ ,
\label{eq:parabolic}
\end{equation}
where $z_{0}$ is the initial position. In the following we will concentrate ourselves on this regime and so we will drop the lower 
scripts l for liquid. It is worth noticing that this solution suffers from two approximations: first, the contact angle is not 
constant; second, inertia effects are not accounted for. 

For channels with radii of millimeter size the weight of the liquid 
contributes to drive the infiltration. This effect is taken into account by adding to the right hand side of Eq.~\ref{eq:washburn} 
the acceleration due to gravity $g=9.8$ [m/s$^2$]. In our description we introduce also the force coming from the column of liquid in 
the reservoir above the channel. Let us assume that the reservoir has sides of length $W_{0}$, $W_{1}$ and $W_{2}$ with 
$r\ll W_{0},W_{1},W_{2}$; $W_{0}$ measures the length along the direction of infiltration $z$. The weight forces deriving from the column 
of liquid above the opening of the channel are described by
\begin{equation*}
\frac{W_{0}W_{1}W_{2}-zr^{2}\pi}{W_{1}W_{2}}r^{2}\pi\rho g\ .
\end{equation*}
The fraction gives the height of the column of liquid above the capillary. In the course of infiltration, the numerator determines
the new volume of liquid in the reservoir. By dividing it by the base area of the reservoir, the new height of the column of liquid
is obtained. Multiplication by the other terms gives the weight of the column of liquid. The above term is added to the right hand
side of Eq.~\ref{eq:capillary}. The differential equation that we want to solve is then given by
\begin{equation}
\ddot{z}=-\frac{(\dot{z})^{2}}{z}+\Big(\frac{2\gamma\cos\theta}{\rho r}+W_{0}g\Big)\frac{1}{z}-\frac{8\mu}{\rho r^{2}}\dot{z}
+\Big(1-\frac{\pi r^{2}}{W_{1}W_{2}}\Big)g\ .
\label{eq:eqdiff}
\end{equation}
For the setup of Fig.~\ref{fig:channel} the last term vanishes.

\section{LB models}\label{sec:model}

LB simulations are particularly suited for our problem \cite{book1,benzi,book2,wolfram,review}. The reason is that interface phenomena can be 
treated exhaustively and efficiently \cite{book1}. More generally, the LB method has an advantage for hydrodynamic studies involving complex 
boundaries. In this work we use a multiphase model in order to describe the process of infiltration \cite{shan1,shan2,succi1,succi3,chibbaro2}. 
With multicomponent models it is possible to reproduce a linear behavior for the penetration dynamics \cite{chibarro,succi2}, but this is not 
necessary since the weight is the dominant force. In any case, it should be kept in mind that multiphase models overestimate the experimental 
density ratio between the gas and liquid phases. Previous works addressed the case of linear infiltration for single channels \cite{supsi1,supsi2}.

Specifically, for a d2q9 lattice the dynamics is obtained by iterating the LB equation
\begin{multline*}
f_{i}(\bm{r}+\bm{e}_{i}\Delta t,t+\Delta t)=
f_{i}(\bm{r},t)-\frac{1}{\tau}\big[f_{i}(\bm{r},t)-
f_{i}^{\mathrm{eq}}(\rho,\bm{u})\big]\\ i=0,\dots,8\ .
\end{multline*}
This equation is based on the BGK approximation \cite{bgk}. The lattice spacing and the time step $\Delta t$ are both assumed to be $1$, without 
loosing generality. The vector $\bm{r}=(x,y)$ points to a lattice site. The allowed velocities are indicated by $\bm{e}_{i}$ and they are 
defined as follows:
\[
\bm{e}_{i}=
\left\{\begin{array}{ll}
(0,0) & i=0\\
(\cos[(i-1)\pi/2],\sin[(i-1)\pi/2]) & i=1-4\\
\sqrt{2}(\cos[(2i-9)\pi/4],\sin[(2i-9)\pi/4]) & i=5-8\ .\\
\end{array}
\right.\]
$f_{i}$ is the distribution function for the particles moving with velocity $\bm{e}_{i}$. In the collision term, the relaxation
time is set to $\tau=1$ [ts]. The relaxation time determines the kinematic viscosity $\nu=(\tau-1/2)/3$. The dynamic viscosity
is given by $\mu=\rho\nu$, where $\rho$ is the density. The equilibrium distribution functions read
\begin{multline}
f^{\mathrm{eq}}_{i}(\rho,\bm{u})=w_{i}\rho(\bm{r},t)\Big[1
+3\bm{e}_{i}\cdot\bm{u}+\frac{9}{2}(\bm{e}_{i}\cdot\bm{u})^{2}
-\frac{3}{2}\bm{u}^{2}\Big]\\ i=0,\cdots,8\ .
\label{eq:feq}
\end{multline}
The weighting factors are $w_{0}=4/9$, $w_{i}=1/9$ for $i=1,2,3,4$ and $w_{i}=1/36$
for $i=5,6,7,8$ \cite{wolfram}. The local density and velocity are computed as follows:
\begin{eqnarray*}
&&\rho(\bm{r},t)=\sum_{i=0}^{8}f_{i}(\bm{r},t)\\ 
&&\bm{u}(\bm{r},t)=\frac{1}{\rho(\bm{r},t)}\sum_{i=0}^{8}\rho(\bm{r},t)f_{i}(\bm{r},t)\bm{e}_{i}\ .
\end{eqnarray*}
The no-slip boundary condition is implemented by using the well-known bounce-back rule. Fluid-fluid interactions give rise to interfaces and 
determine the surface tension $\gamma$. The cohesive forces are introduced by means of the function \cite{shan1,shan2}
\begin{equation*}
\bm{F}_{\mathrm{c}}(\bm{r})=-G_{\mathrm{c}}\psi(\bm{r})\sum_{i=1}^{8}w_{i}\psi(\bm{r}+\bm{e}_{i}\Delta t)\bm{e}_{i}\ .
\end{equation*}
The parameter $G_{\mathrm{c}}$ controls the interaction strength. The function $\psi$ is defined as $\psi(\bm{r})=\psi_{0}e^{-\rho_{0}/\rho(\bm{r})}$. 
$\psi_{0}$ and $\rho_{0}$ are free parameters. The pressure is given by \cite{book1}
\begin{equation*}
P(\bm{r})=\frac{\rho(\bm{r})}{3}+\frac{G_{\mathrm{c}}}{6}\psi^{2}(\bm{r})\ .
\end{equation*}
The adhesive forces between the fluid and solid phases determine the contact 
angle. They are computed with
\begin{equation*}
\bm{F}_{\mathrm{ads}}(\bm{r})=-G_{\mathrm{ads}}\psi(\bm{r})\sum_{i=1}^{8}w_{i}s(\bm{r}+\bm{e}_{i}\Delta t)\bm{e}_{i}\ .
\end{equation*}
The parameter $G_{\mathrm{ads}}$ sets the interaction strength. The function $\psi$ is the same as for cohesive forces. The function
$s(x,y)$ takes on the value $1$ if the site is occupied by the solid phase, otherwise it is equal to $0$. The gravitational force
is introduced by imposing the constant acceleration $\bm{g}$. Finally, the contribution of all the forces is taken into 
account by using in Eq.~\ref{eq:feq} the velocity $\bm{u}+\tau(\bm{F}_{\mathrm{c}}+\bm{F}_{\mathrm{ads}}+\rho\bm{g})/\rho$.

Solute transport takes place on a d2q4 lattice \cite{d2q4}, without diagonal velocities. As for fluid flow, the evolution of the
distribution functions is given by
\cite{d2q4,reaction_pre,dissolution,crystal,snow}
\begin{multline*}
g_{i}(\bm{r}+\bm{e}_{i}\Delta t,t+\Delta t)=g_{i}(\bm{r},t)
-\frac{1}{\tau_{\mathrm{s}}}\big[g_{i}(\bm{r},t)-g_{i}^{\mathrm{eq}}(C,\bm{u})\big]\\
i=1,\cdots,4\ .
\end{multline*}
$\tau_{\mathrm{s}}=1$ [ts] is the relaxation time for solute transport. The diffusion coefficient is given by 
$D=(2\tau_{\mathrm{s}}-1)/4$ \cite{d2q4}. The equilibrium distribution functions become \cite{d2q4}
\begin{equation}
g_{i}^{\mathrm{eq}}(C,\bm{u})=\frac{1}{4}C\big[1+2\bm{e}_{i}\cdot\bm{u}\big]
\quad i=1,\cdots,4\ .
\label{eq:geq}
\end{equation}
$\bm{u}$ is the local solvent velocity; the solute concentration is defined as $C=\sum_{i}g_{i}$. In order to limit diffusion
at the meniscus of the invading front we create an interface by using the function
\begin{equation}
\bm{F}_{\mathrm{s}}(\bm{r})=-G_{\mathrm{s}}\varphi(\bm{r})\sum_{i=1}^{4}\varphi(\bm{r}+\bm{e}_{i}\Delta t)\bm{e}_{i}\ .
\label{eq:fs}
\end{equation}
The parameter $G_{\mathrm{s}}$ is used in order to adjust the intensity. The function $\varphi$ is defined as 
$\varphi(\bm{r})=\varphi_{0}e^{-C_{0}/C(\bm{r})}$; $\varphi_{0}$ and $C_{0}$ are arbitrary constants. In analogy with fluid flow,
we add to the velocity in Eq.~\ref{eq:geq} the term $\tau\bm{F}_{\mathrm{s}}/C$. The surface reaction at the boundaries is described by the 
macroscopic equation \cite{d2q4,reaction_pre,dissolution,crystal,snow}
\begin{equation*}
D\frac{\partial C}{\partial n}=k_{\mathrm{r}}\big(C-C_{\mathrm{s}}\big)\ .
\end{equation*}
$k_{\mathrm{r}}$ is the reaction-rate constant and $C_{\mathrm{s}}$ is the saturated concentration. In the LB formalism, this
condition takes the form \cite{d2q4}
\begin{equation*}
C=\frac{2g_{\mathrm{\chi(i)}}+k_{\mathrm{r}}C_{\mathrm{s}}}{k_{\mathrm{r}}+1/2}
\quad\text{and}\quad g_{i}=-g_{\chi(i)}+\frac{1}{2}C\ .
\end{equation*}
$i$ is the index of the distribution functions that remain undetermined after the streaming step; $\chi(i)$ is the index for the
opposite velocity, leaving the fluid phase. Mass deposits on the solid surface. Up to a constant, the mass that 
accumulates on the solid nodes belonging to the solid-liquid interface is obtained by iterating the equation \cite{crystal,snow}
\begin{equation*}
b(\bm{r},t+\Delta t)=b(\bm{r},t)+k_{\mathrm{r}}(C-C_{\mathrm{s}})\ .
\end{equation*}
The initial mass is $b_{0}$. When the mass $b$ exceeds the threshold value $b_{\mathrm{max}}$, the surface grows. This means that one
of the direct neighbors in the fluid phase becomes a solid node. This node is chosen with equal probability among the available 
possibilities. The mass of the solid node that triggered surface growth is set to zero.

Hereafter, the output of the simulations will be expressed in model units. The symbols for the basic quantities of length, mass and time
are lu, mu and ts, respectively. In this notation, ts stands for timesteps. In the sequel, we will try to compare simulations with
experiments by means of dimensionless numbers like the capillary number, the Damkohler number \cite{crystal} and other numbers involving
the surface reaction. This procedure guarantees the equivalence of the conditions determining the hydrodynamic behavior \cite{landau}.

\begin{figure}[t]
\includegraphics[width=8.4cm]{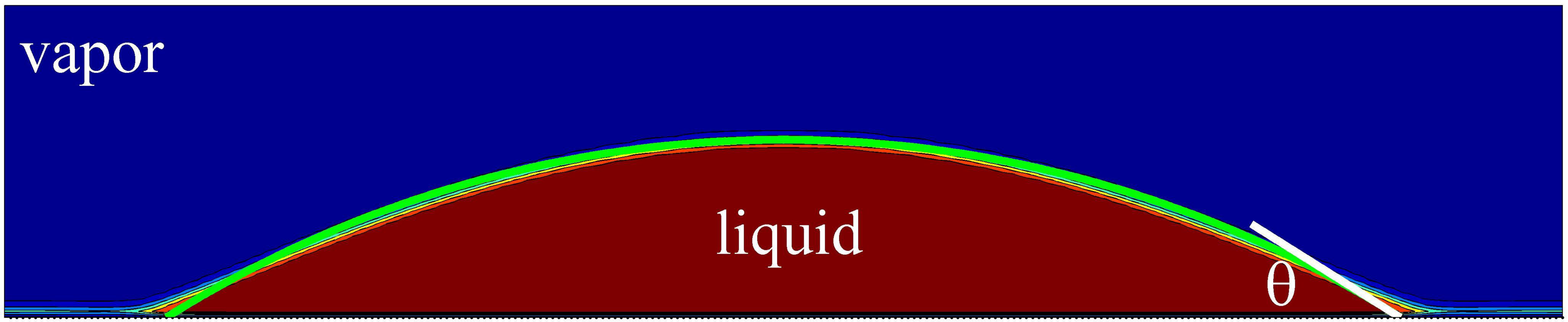}
\caption{\label{fig:droplet}
Sessile droplet for the determination of the contact angle $\theta$ using LB simulations. The white straight line is the tangent at the
contact line. The green curve represents the circle fitting to the interface.}
\end{figure}
\begin{figure}[t]
\includegraphics[width=8.4cm]{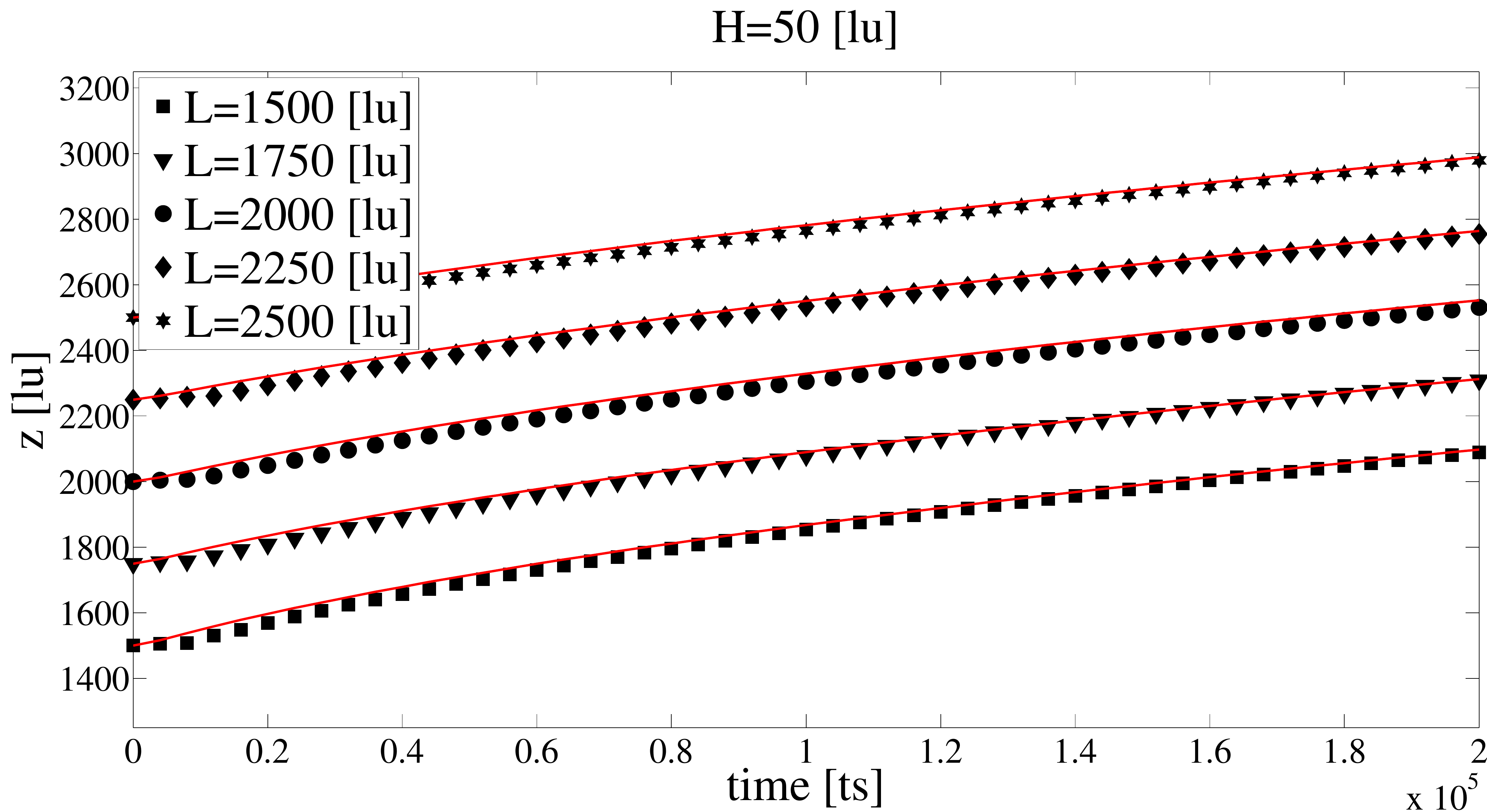}
\caption{\label{fig:front}
LB simulations for capillary infiltration driven only by adhesive forces. Position of the invading front as time goes on. Points 
for simulation results; solid lines for the theoretical predictions of Eq.~\ref{eq:washburn} in 2D. $L$ is the
length of the capillary and $H$ its height.}
\end{figure}
\begin{figure}[t]
\includegraphics[width=8.4cm]{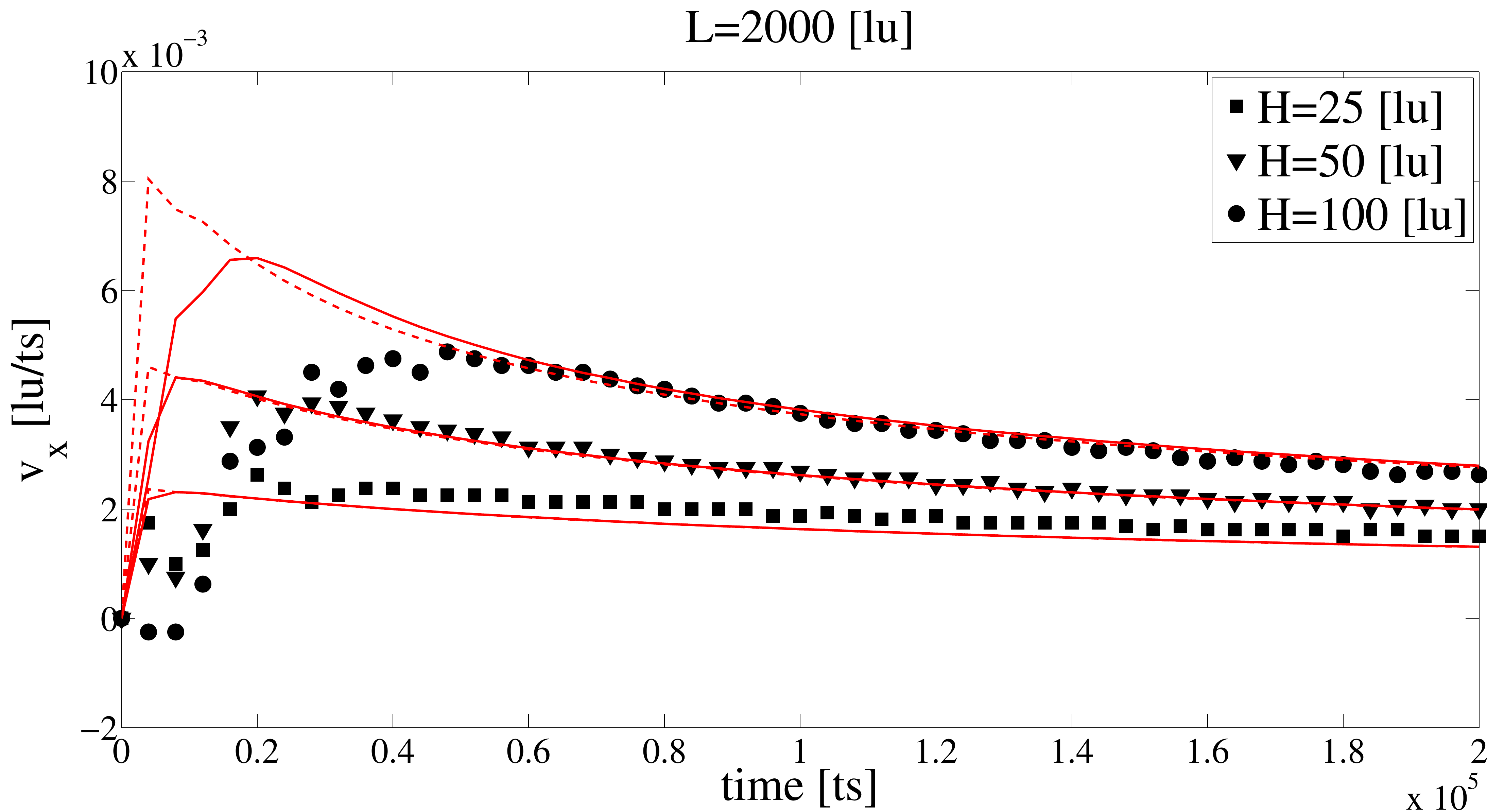}
\caption{\label{fig:vx}
LB simulations for infiltration driven only by capillary forces. Velocity of the invading front as a function of time.
The points represent the results from simulations. The solid line is obtained by solving Eq.~\ref{eq:washburn}
in the 2D case. The dashed lines correspond to Eq.~\ref{eq:parabolic} in 2D. $L$ denotes the length of the
channel and $H$ the height.}
\end{figure}
\begin{figure}[t]
\includegraphics[width=8.4cm]{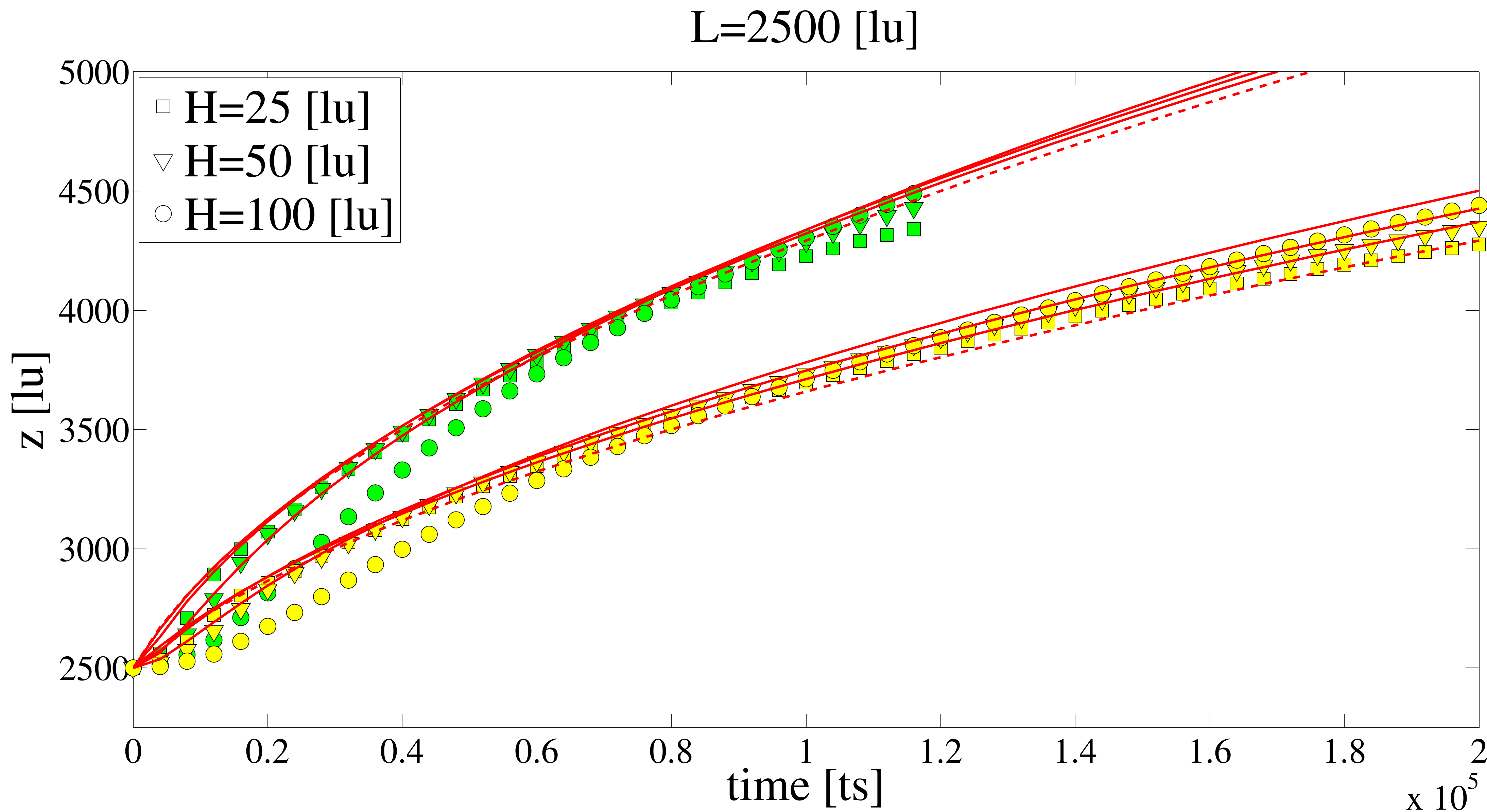}
\caption{\label{fig:g}
Front displacement in the course of time for infiltration with a constant acceleration and inert boundaries. Points 
represent the results of LB simulations. For the data in green, the condition $gH^{2}=2\cdot10^{-2}$ [lu$^{3}$/ts$^{2}$] 
holds, while the data in yellow satisfy $gH^{2}=10^{-2}$ [lu$^{3}$/ts$^{2}$]. The red solid lines are solutions to 
Eq.~\ref{eq:eqdiff} in 2D. The dashed lines correspond to the predictions of Eq.~\ref{eq:z_w} in 2D.}
\end{figure}
\begin{figure*}[t]
\includegraphics[width=8.4cm]{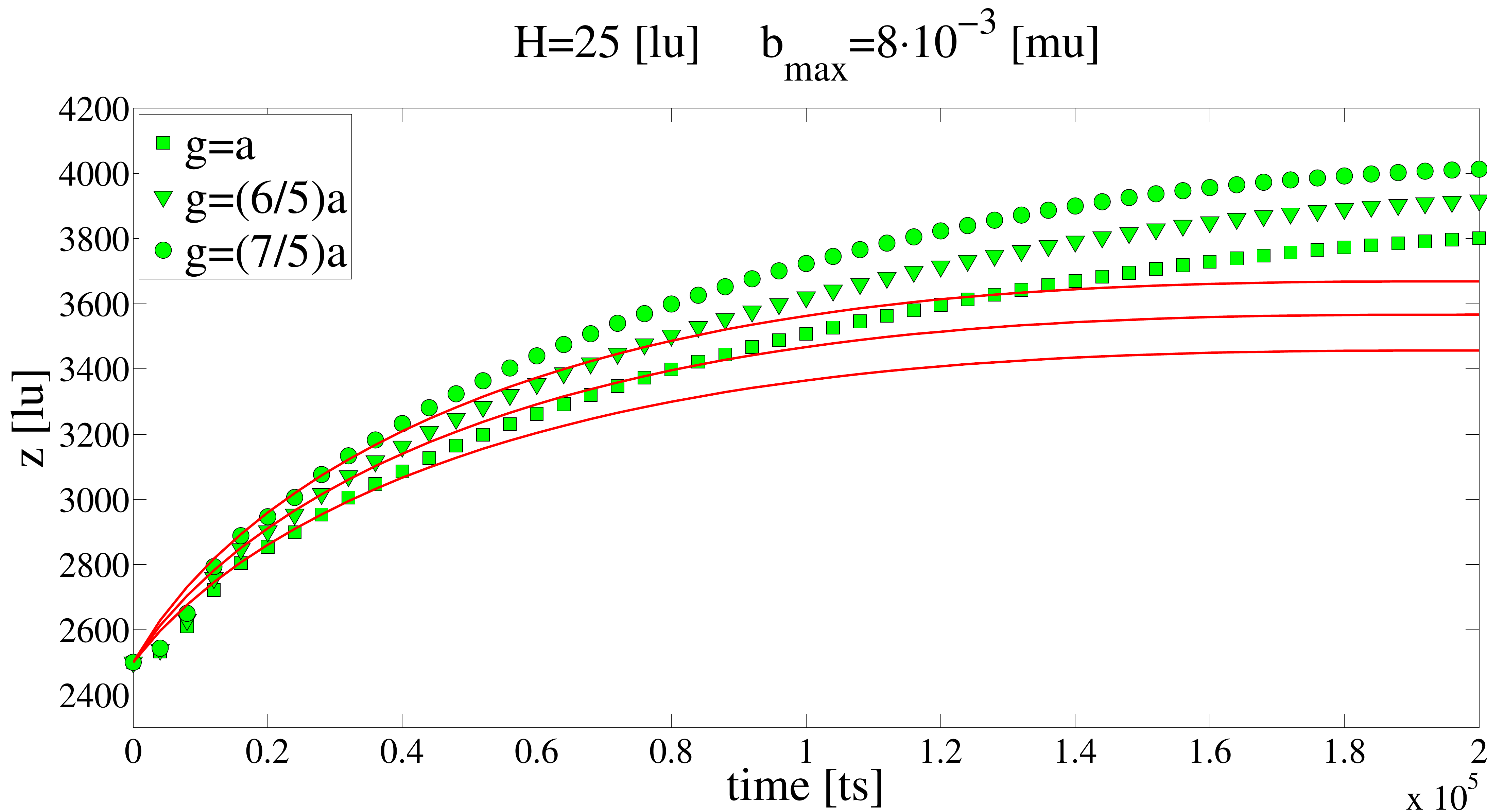}
\includegraphics[width=8.4cm]{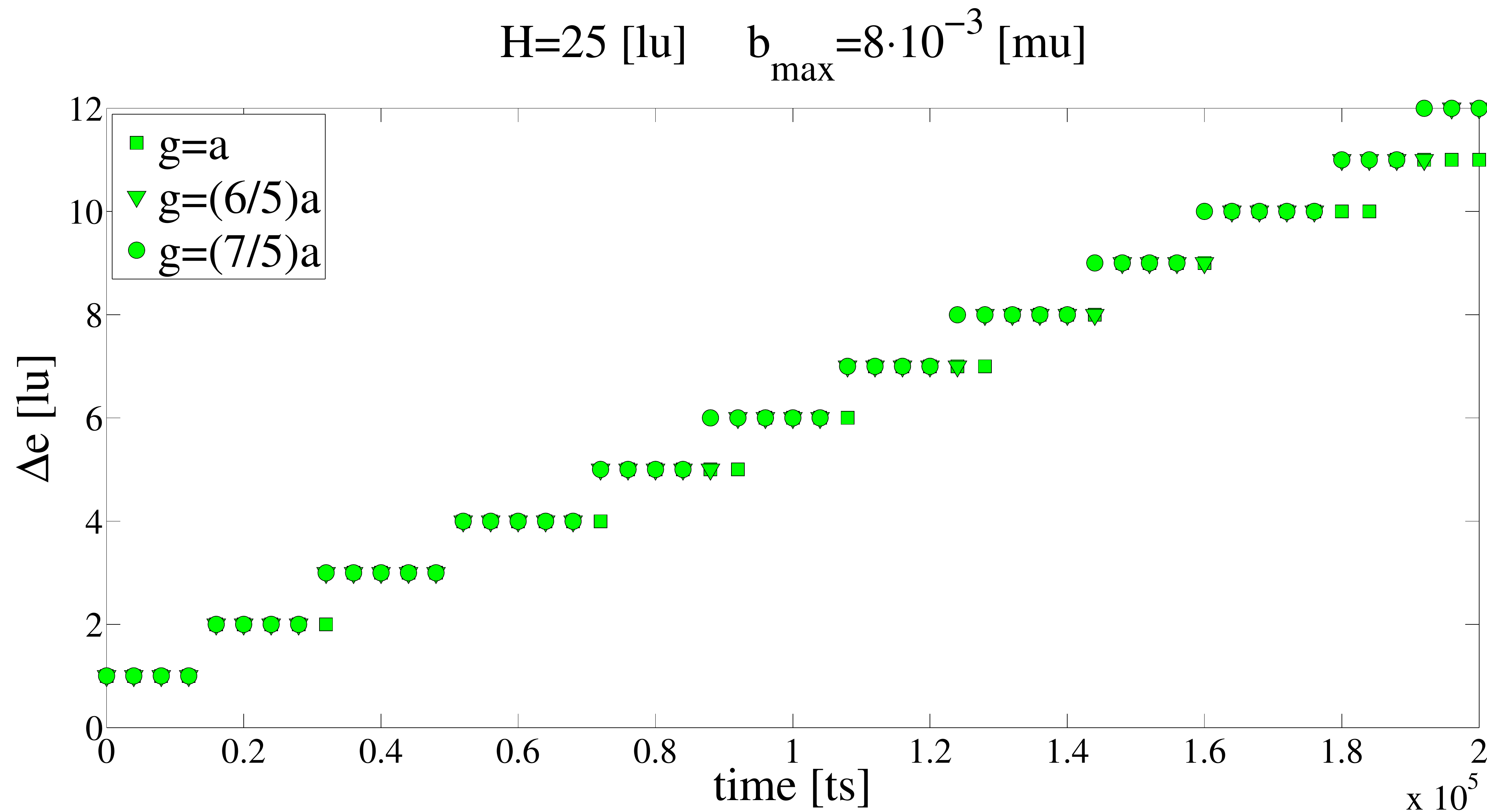}\\
\includegraphics[width=8.4cm]{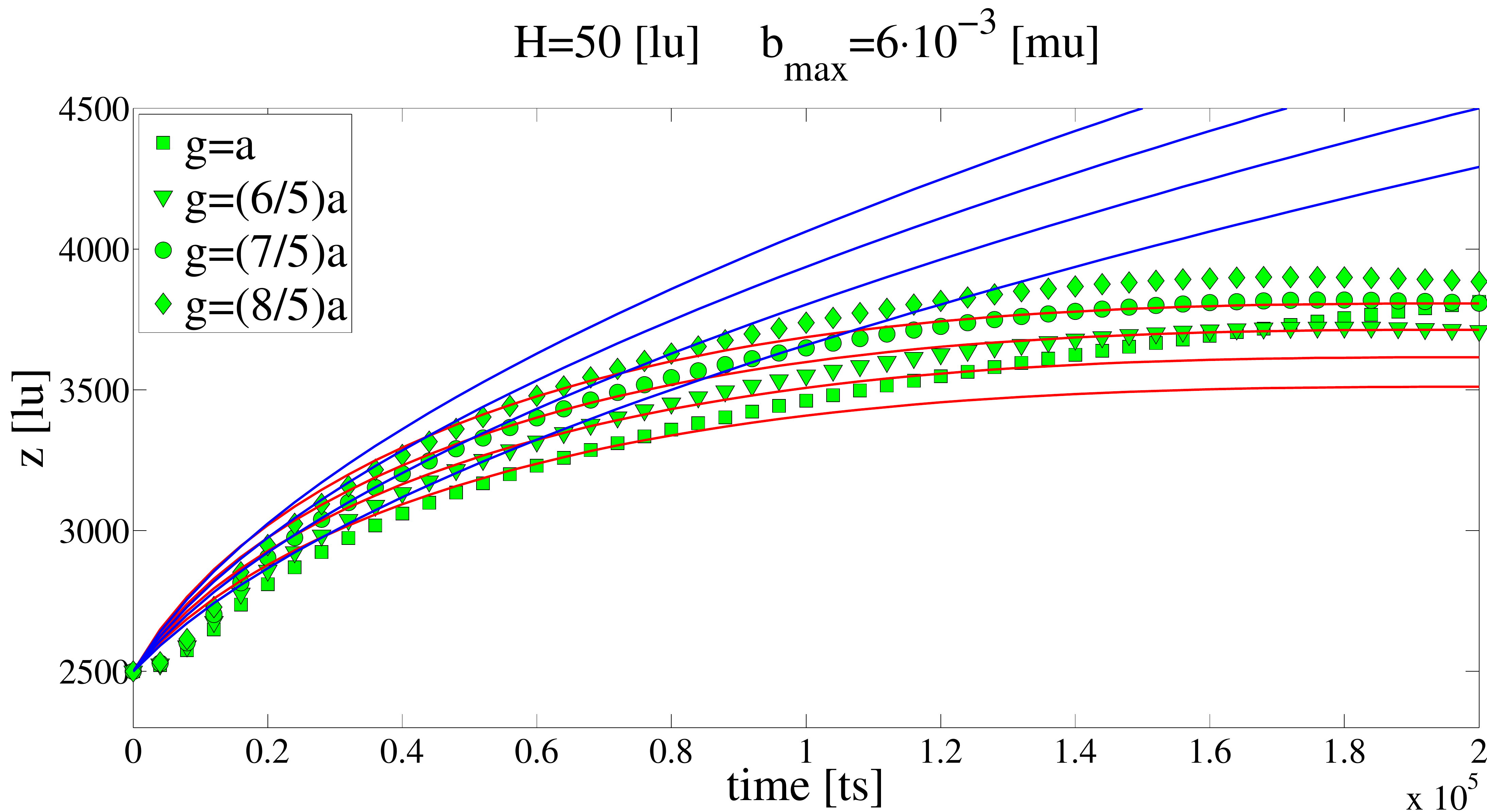}
\includegraphics[width=8.4cm]{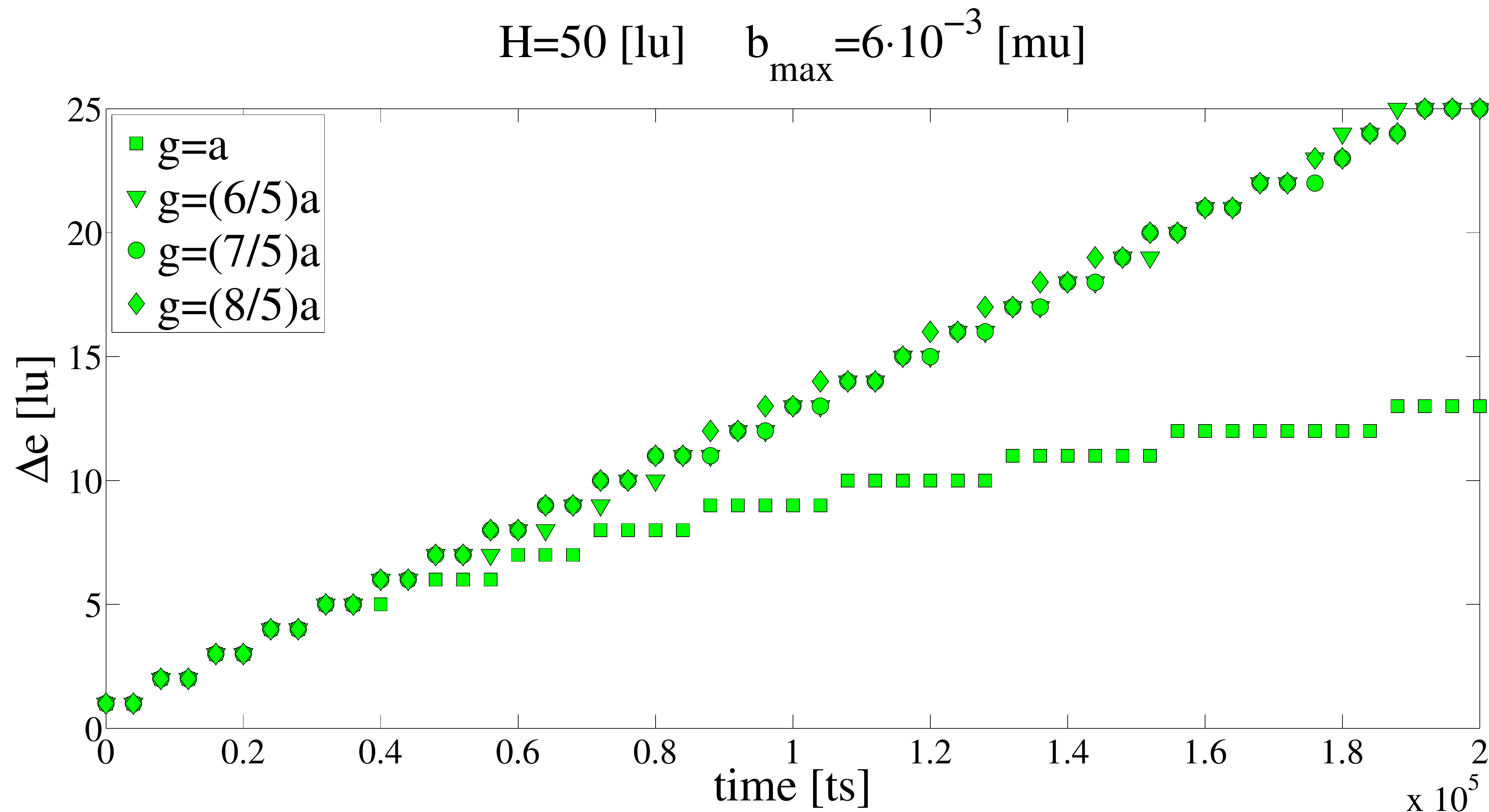}\\
\includegraphics[width=8.4cm]{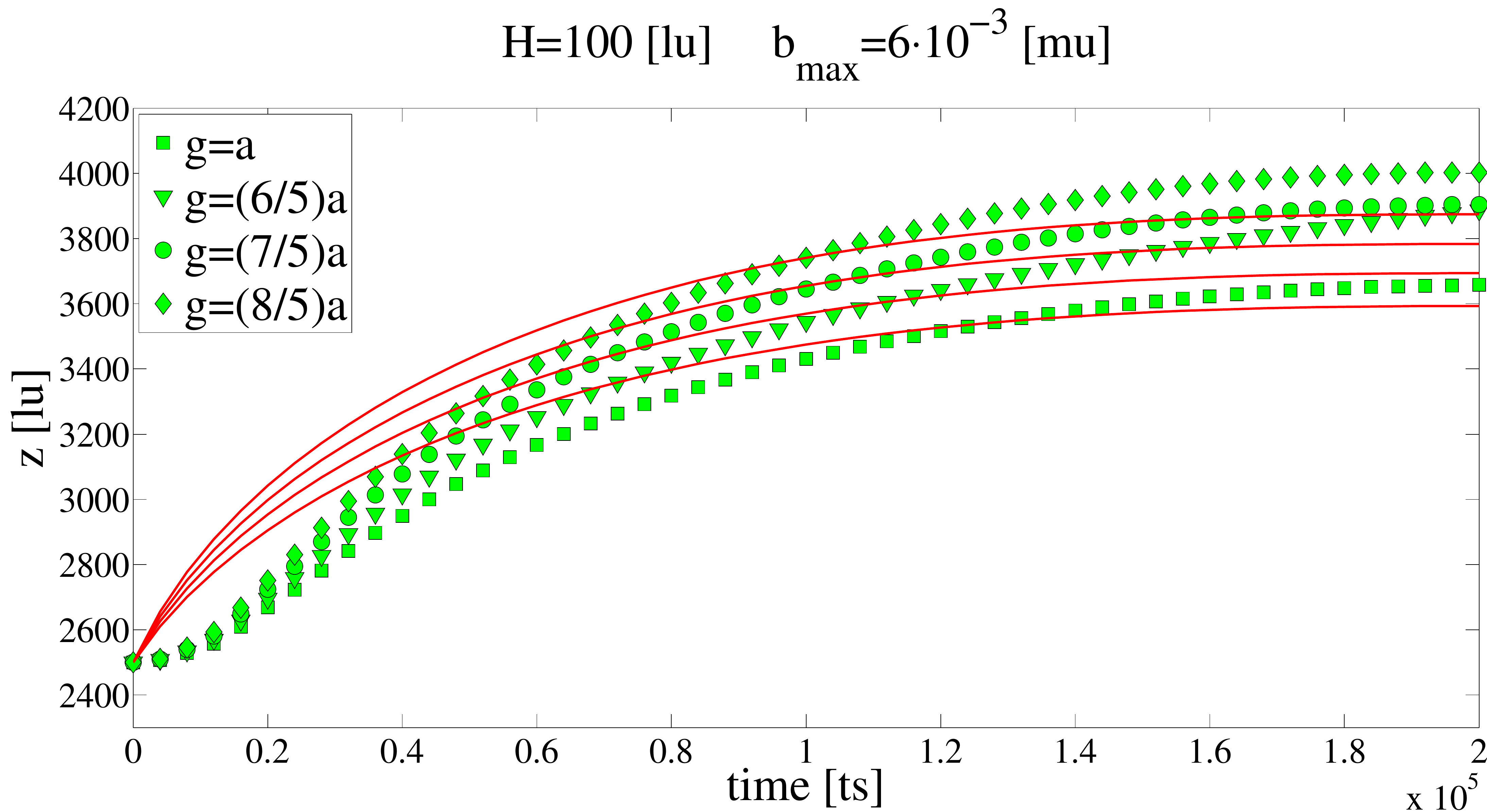}
\includegraphics[width=8.4cm]{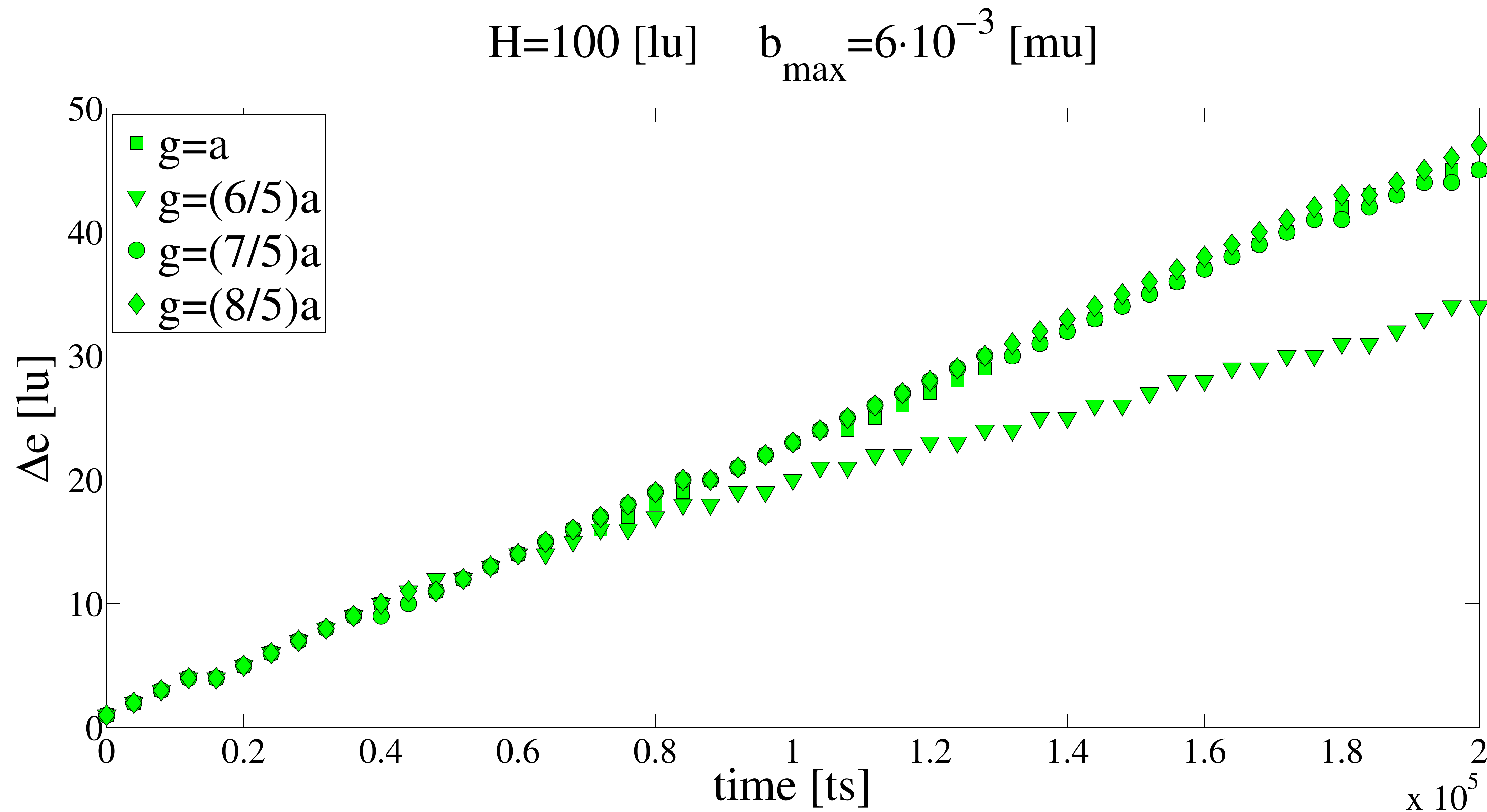}
\caption{\label{fig:reaction}
Results of simulations in the presence of surface reaction. The length of the capillary is $L=2500$ [lu]. The reaction-rate 
constant is chosen so that $k=(1/2)H/N_{\mathrm{t}}$, where $N_{\mathrm{t}}=200000$ [ts] is the number of timesteps. The first value 
for the acceleration is given by $a=10^{-2}/H^{2}$, as in Fig.~\ref{fig:g}. Left: Front displacement as time evolves. Points are 
associated with simulation results. The red solid lines correspond to the theoretical predictions of Eq.~\ref{eq:z_theta} in 2D. 
The blue lines represent the predictions of Eq.~\ref{eq:z_w} in 2D. Right: Maximal width of the growing surface in the course of 
time. Points are used for results from simulations.}
\end{figure*}
\begin{figure}[t]
\includegraphics[width=8.4cm]{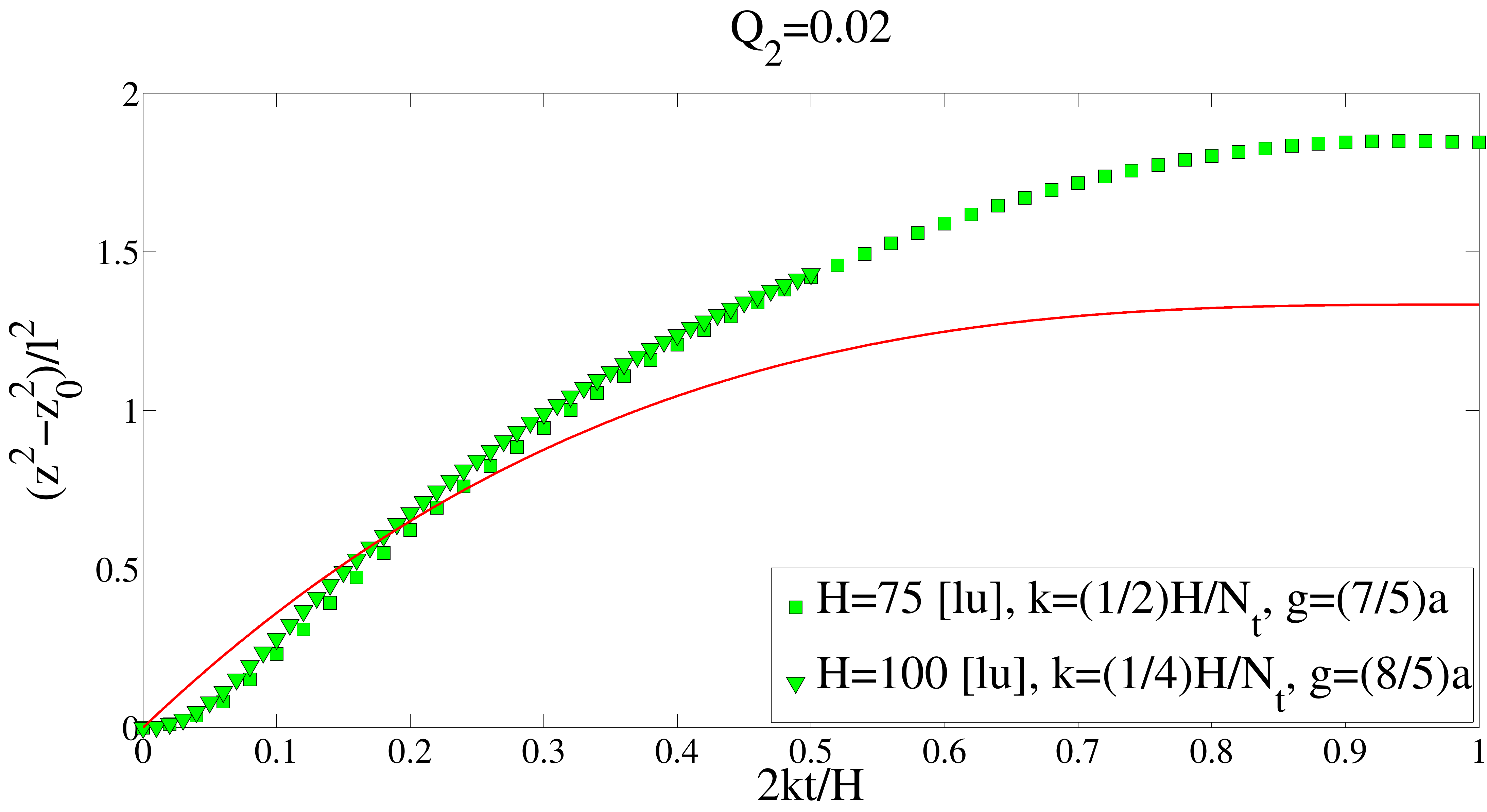}
\caption{\label{fig:equivalence}
Equivalence of systems based on the relative strength between reaction and gravity, determined by the characteristic
number $Q_{2}=vk/(rg)$. The results are plotted for the 2D case according to the representation suggested by 
Eq.~\ref{eq:dimless}. In this specific case, the quantity $\mathit{l}$ is given by $l=\sqrt{gLr_{0}^{3}/k}$. For the 
acceleration we use $a=10^{-2}/H^{2}$; $N_{\mathrm{t}}=200000$ [ts] is the total number of timesteps}
\end{figure}


\section{Analysis}

The Bond number compares gravity and surface tension: $Bo=r^{2}\rho g/\gamma$. The capillary number compares the effects of viscosity and 
surface tension: $Ca=v\mu/\gamma$, where $v$ is an estimate of the characteristic velocity. In order to compare gravity and viscosity, one
has to consider the number
\begin{equation}
Q_{1}=\frac{Bo}{Ca}=\frac{\rho r^{2}g}{v\mu}\ .
\label{eq:q1}
\end{equation}
We assume that this dimensionless number characterizes our systems during infiltration since the gravitational forces find the resistance of
viscous forces. This means that the resistance of viscosity has a stronger effect than the driving force due to adhesive forces. The results
of direct simulations with Eq.~\ref{eq:eqdiff} indicate that the penetration depth with gravity is always smaller than in the case of free fall.
As a consequence, if gravitational forces are dominant, the typical time scale is set by $t_{g}=v/g$. Note that the velocity $v$ takes into
account also the effects of viscosity. Furthermore, under these conditions the centerline position of the invading front is given by
\begin{equation}
z=\frac{\rho r^{2}g}{8\mu}t+z_{0}\ .
\label{eq:z1}
\end{equation}
Numerical solutions to Eq.~\ref{eq:eqdiff} with gravity for oil \cite{oil} return results in good agreement with the above equation. From 
Eq.~\ref{eq:q1} we obtain $Q_{1}=8$ for any $g$ realizing the regime of interest.

The time scale associated with the reactivity is given by $t_{k}=r/k$, where $k$ is the reaction-rate constant. Roughly speaking, this is the
time when pore closure occurs. The relative effects between the reactivity and gravitational forces are thus determined by the number
\begin{equation*}
Q_{2}=\frac{vk}{rg}\ .
\end{equation*}
Higher values of $Q_{2}$ mean that reactivity becomes faster than mechanics, i.e.~the hydrodynamics mainly determined by gravitational forces. 
For completeness, $Q_{3}=\rho r k/\mu$ compares the reaction with the viscosity and $Q_{4}=\rho rvk/\gamma$ compares the reaction and surface tension.
The dimensionless numbers are compared in Tab.~\ref{tab:experimental} for a typical capillary embedded in a porous medium and a millimeter-sized
capillary. For $r=10$ [$\mu$m] it is evident that capillary forces drive the infiltration. In general, it is assumed that high values
for the capillary number satisfy $Ca>0.01$. $Q_{1}$ is high but the viscous resistance is in response to capillary forces since $Ca^{-1}$
is much higher. The same conclusions can be drawn by analyzing the other dimensionless numbers.
In passing to $r=0.5$ [mm], gravity gains three orders of magnitude with respect to capillarity. As a test, if we solve Eq.~\ref{eq:eqdiff},
we find that the term of weight forces is $15$ times larger than that due to surface tension already after $2$ [sec]. Furthermore, the accordance
with Eq.~\ref{eq:z1} is good. 
Another dimensionless number that is worth mentioning is the Damkohler number. It is defined as $Da=kH/D$, where $H$ is a characteristic distance
of the systems. For example, in a purely diffusive regime, this number determines the structure of the growing surface: small values for compact
structures and large values for dendritic morphologies \cite{crystal}.
\begin{table*}[t]
\begin{ruledtabular}
\begin{tabular}{lcccccc}
$r$ & $Q_{1}$ & $Q_{2}$ & $Q_{3}$ & $Q_{4}$ & $Bo$ & $Ca$\\
\hline
$10$  [$\mu$m] & $2.6\cdot 10^{3}$ & $4.1\cdot 10^{-10}$ & $1.1\cdot 10^{-6}$ & $1.2\cdot 10^{-15}$ & $2.9\cdot 10^{-6}$ & $1.1\cdot 10^{-9}$\\
$0.5$ [mm]     & $8$              & $6.7\cdot 10^{-6}$  & $5.4\cdot 10^{-5}$ & $4.8\cdot 10^{-8}$  & $7.2\cdot 10^{-3}$ & $9\cdot 10^{-4}$\\
\end{tabular}
\end{ruledtabular}
\caption{\label{tab:experimental}
Characteristic numbers for capillaries of different size infiltrated with molten Si \cite{einset1, messner}. $Q_{1}$ compares the relative importance of gravity and viscosity,
$Q_{2}$ of reactivity and gravity, $Q_{3}$ of reactivity and viscosity, $Q_{4}$ of reactivity and surface tension; $Bo$ compares gravity and viscosity 
(Bond number) and $Ca$ compares viscosity and surface tension (capillary number). The reaction-rate constant is set to $k=4\cdot10^{-8}$ [m/s]
\cite{messner}. The infiltration velocity for the narrow channel is approximated by $v=1$ [$\mu$m/s] \cite{nisi,nisi2,htc09}. For  $r=0.5$ [mm],
the velocity is estimated using Eq.~\ref{eq:z1}. The surface tension of Si is $\gamma=0.86$ [N/m], the dynamic viscosity 
$\mu=0.94\cdot10^{-3}$ [N$\cdot$s/m$^{2}$] and the density $\rho=2.53\cdot10^{3}$ [kg/m$^{3}$] \cite{einset1}.}
\end{table*}

For a shrinking radius the infiltration kinetics becomes
\begin{equation*}
z=\frac{\rho g}{24\mu k}\big[r_{0}^{3}-(r_{0}-kt)^{3}\big]+z_{0}\ ,
\end{equation*}
where $z(t=0)=z_{0}$. This equation is obtained by assuming that in Eq.~\ref{eq:eqdiff} gravity and viscosity are the dominant forces,
as for Eq.~\ref{eq:z1}. Furthermore, here reactivity is implemented by imposing that the initial radius $r_{0}$ shrinks according to the 
condition $r=r_{0}-kt$, where $r_{0}$ is the initial value \cite{messner}. The maximal penetration depth is attained at the time $t=r_{0}/k$, 
leading to 
\begin{equation*}
z_{\mathrm{max}}=\frac{\rho g r_{0}^{3}}{24\mu k}+z_{0}\ .
\end{equation*}
A simple calculation indicates that without reaction the infiltration depth would be $3z_{\mathrm{max}}$ during the same time lag with 
$z_{0}=0$ [lu].

If we take into account the weight of the liquid in the reservoir as in Eq.~\ref{eq:eqdiff}, for the initial condition of 
LB simulations Eq.~\ref{eq:z1} becomes
\begin{equation}
z=\sqrt{\frac{\rho r^{2} L g}{4\mu}t+z_{0}^{2}}\ ,
\label{eq:z_w}
\end{equation}
with $W_{0}=L$ the length of the channel. The length of the simulation domain is given by $N_{x}=2L$. For a shrinking radius we obtain
\begin{equation}
z=\sqrt{\frac{\rho L g}{12\mu k}\big[r_{0}^{3}-(r_{0}-kt)^{3}\big]+z_{0}^{2}}\ .
\label{eq:z_wk}
\end{equation}
The maximal penetration depth is in this case
\begin{equation*}
z_{\mathrm{max}}=\sqrt{\frac{\rho L g r_{0}^{3}}{12\mu k}+z_{0}^{2}}\ .
\end{equation*}
During the same time interval, without reaction the infiltration depth would be $\sqrt{3}z_{\mathrm{max}}$
by assuming that $z_{0}=0$ [lu]. The above results can be useful in order to obtain estimates for the
characterization of experiments. For the sake of completeness, if capillary forces remain significant
it is obtained
\begin{equation}
z=\sqrt{\frac{\rho L g}{12\mu k}\big[r_{0}^{3}-(r_{0}-kt)^{3}\big]+\frac{\gamma\cos\theta}{4\mu k}
\big[r_{0}^{2}-(r_{0}-kt)^{2}\big]+z_{0}^{2}}\ .
\label{eq:z_theta}
\end{equation}
The maximal penetration depth becomes
\begin{equation*}
z_{\mathrm{max}}=\sqrt{\frac{\rho L g r_{0}^{3}}{12\mu k}+\frac{\gamma r_{0}^{2}\cos\theta}{4\mu k}+z_{0}^{2}}\ .
\end{equation*}
Equation \ref{eq:z_wk} can be rewritten in dimensionless form as
\begin{equation}
\tilde{z}^{2}-\tilde{z}_{0}^{2}=\frac{1}{12}\big[1-(1-\tilde{t})^{3}\big]\ ,
\label{eq:dimless}
\end{equation}
where $\tilde{t}=kt/r_{0}$ and $\tilde{z}=z/\mathit{l}$ with $\mathit{l}=\sqrt{\rho Lgr_{0}^{3}/(\mu k)}$.
Instead, if capillary forces are the dominant ones, Eq.~\ref{eq:z_theta} can be recast into dimensionless 
form as
\begin{equation*}
\tilde{z}^{2}-\tilde{z}_{0}^{2}=\frac{\cos\theta}{4}\big[1-(1-\tilde{t})^{2}\big]\ ,
\end{equation*}
where $\tilde{t}=kt/r_{0}$ and $\tilde{z}=z/\mathit{l}$ with $\mathit{l}=r_{0}\sqrt{\gamma /(\mu k)}$.


\section{Simulation results}

\subsection{Surface tension}

To start with, we determine the surface tension $\gamma$ using Laplace law. This law states that, for a liquid bubble immersed 
in its vapor phase, the pressure drop across the interface is proportional to $\gamma$. In 2D, Laplace law reads 
$\Delta P=P_{\mathrm{in}}-P_{\mathrm{out}}=\gamma/r$. For cohesive forces $\bm{F}_{\mathrm{c}}$, the potential parameters are set to 
$G_{\mathrm{c}}=-140$ [mu$\cdot$lu/ts$^{2}$], $\psi_{0}=4$ and $\rho_{0}=200$ [mu/lu$^{2}$]. It is found that the surface tension 
is $\gamma=29.91$ [lu$\cdot$mu/ts$^{2}$]. The densities for the liquid and the gas are $\rho_{\mathrm{l}}=700$ [mu/lu$^{2}$] and
$\rho_{\mathrm{g}}=70$ [mu/lu$^{2}$]. The density ratio is thus $\rho_{\mathrm{l}}/\rho_{\mathrm{g}}=10$.

\subsection{Contact angle}

The equilibrium contact angle $\theta$ is obtained by using the method of the sessile droplet. A droplet of liquid is put
in contact with the surface. The droplet spreads in order to reach the equilibrium state consisting of a spherical cap. The
tangent at the contact line, where the three phases meet, defines the contact angle (see Fig.~\ref{fig:droplet}). For adhesive 
forces $\bm{F}_{\mathrm{ads}}$, the potential parameters are given by  $G_{\mathrm{ads}}=-380$ [lu/ts$^{2}$], $\psi_{0}=4$ and 
$\rho_{0}=200$ [mu/lu$^{2}$]. The outcome for the equilibrium contact angle is $\theta=32^{\circ}$. The experimental value 
for the contact angle is $30^{\circ}$ \cite{nisi,nisi2}.

\subsection{Washburn infiltration}

In experiments based on pure Si and some alloys with porous C, the infiltration process exhibits a linear time dependence
for the displacement of the invading front \cite{nisi,nisi2,htc09}. The breakdown of the ordinary Washburn behavior of 
Eq.~\ref{eq:parabolic} is ascribed to the effects due to the reaction between Si and C to form SiC \cite{nisi,nisi2,htc09}. 
Since gravity is expected to be the dominant force, in this study we use models reproducing the usual parabolic time 
dependence \cite{shan1,shan2,succi1,succi3,chibbaro2}. In our systems, the liquid to vapor density ratio is $10$, so quite 
high. First, we thus evaluate the implications of this approximation. 
Simulations for the process of infiltration are performed with systems similar to that shown in Fig.~\ref{fig:channel}.
Figure \ref{fig:front} shows the dependence of the 
results on the length of the capillary $L$ for a given height $H$. It turns out that, as the height increases, the departure 
from the predictions of Eq.~\ref{eq:washburn} is more pronounced. The dependence on the length of the channel is less marked. 
It is important to recall that what characterizes the hydrodynamic behavior is the velocity. Figure \ref{fig:vx} compares the 
velocity to the predictions of the models. Also for $H=100$ [lu] the velocity is almost equal to that obtained with a vanishing 
density for the gas phase relatively soon. This implies that the simulated systems reproduce correctly the desired Reynolds 
number and capillary number of the analytical and numerical solutions. As a result, in the sequel simulations of validation 
will be carried out with capillaries of height at most of $100$ [lu]. 

\subsection{Capillary infiltration with gravity}

We consider capillary infiltration in the presence of a constant acceleration mimicking gravity. Figure \ref{fig:g} shows the 
results for LB simulations. The two groups of data are characterized by different values for the product $gH^{2}$. As predicted 
by Eq.~\ref{eq:q1}, we see that within a group of results the behavior of the systems is comparable (cf.~Eq.~\ref{eq:z_w}), with
the exception for the initial stages of the dynamics. The deviations are more marked for increasing width of the channels and for
increasing acceleration. Under these conditions the viscous and inertia resistance of the gas phase results to be stronger. The
main effect is that at the beginning the process is particularly slow. As a result, we consider the present systems reliable for
the investigation of the properties of surface growth. Table \ref{tab:numbers0} reports the characteristic numbers of relevance
for these settings. As indicated by the simulation results, these values are associated with a dominant force given by gravity.

\begin{table}[t]
\begin{ruledtabular}
\begin{tabular}{lccc}
$H$ [lu] & $Q_{1}$ & $Bo$ & $Ca$\\
\hline
$25$     & $1.9$ & $5.7\cdot 10^{-2}$ & $2.9\cdot 10^{-2}$\\
$50$     & $1.8$ & $5.8\cdot 10^{-2}$ & $3.2\cdot 10^{-2}$\\
$100$    & $1.6$ & $5.8\cdot 10^{-2}$ & $3.7\cdot 10^{-2}$\\
\end{tabular}
\end{ruledtabular}
\caption{\label{tab:numbers0}
Characteristic numbers for infiltration driven by gravity without surface reaction (see Fig.~\ref{fig:g}). The constant acceleration
is $g=10^{-2}/H^{2}$. $H$ is the width of the channel. $Q_{1}$ compares gravitational and viscous forces. $Bo$ compares gravity and surface
tension (Bond number), while $Ca$ compares viscosity and surface tension (capillary number). The quantities that can vary 
in the course of time are averaged over a time interval centered around $N_{\mathrm{t}}/2=100000$ [ts].}
\end{table}
\begin{table}[t]
\begin{ruledtabular}
\begin{tabular}{lcccccc}
$H$ [lu] & $Q_{1}$ & $Q_{2}$ & $Q_{3}$ & $Q_{4}$ & $Bo$ & $Ca$\\
\hline
$25$  & $0.87$ & $2.9\cdot 10^{-3}$ & $2.5\cdot 10^{-3}$ & $5.4\cdot 10^{-5}$ & $1.9\cdot 10^{-2}$ & $2.2\cdot 10^{-2}$\\
$50$  & $0.96$ & $1.1\cdot 10^{-2}$ & $1.0\cdot 10^{-2}$ & $2.2\cdot 10^{-4}$ & $2.1\cdot 10^{-2}$ & $2.2\cdot 10^{-2}$\\
$100$ & $0.93$ & $4.6\cdot 10^{-2}$ & $4.3\cdot 10^{-2}$ & $1.1\cdot 10^{-3}$ & $2.5\cdot 10^{-2}$ & $2.6\cdot 10^{-2}$\\
\end{tabular}
\end{ruledtabular}
\caption{\label{tab:numbers}
Characteristic numbers for simulations with constant acceleration and surface reaction (see Fig.~\ref{fig:reaction}). $H$ is the 
initial width of the channel. $Q_{1}$ compares gravitational and viscous forces. $Q_{2}$, $Q_{3}$ and $Q_{4}$ compare the effects 
of reaction to gravitational, viscous and capillary forces. $Bo$ is the Bond number comparing gravity and surface tension; 
$Ca$ is the capillary number comparing viscosity and surface tension. The results for a given initial height are averaged 
over the different applied accelerations (see Fig.~\ref{fig:reaction}). The quantities that can vary as time evolves 
are averaged inside a window centered around the time $N_{\mathrm{t}}/2=100000$ [ts].}
\end{table}
\begin{table}[b]
\begin{ruledtabular}
\begin{tabular}{lcccccc}
$H$ [lu] & $Q_{1}$ & $Q_{2}$ & $Q_{3}$ & $Q_{4}$ & $Bo$ & $Ca$\\
\hline
$25$  & $1.3$ & $1.3\cdot 10^{-3}$ & $1.7\cdot 10^{-3}$ & $4.6\cdot 10^{-5}$ & $3.6\cdot 10^{-2}$ & $2.7\cdot 10^{-2}$\\
$50$  & $1.4$ & $4.9\cdot 10^{-3}$ & $7.0\cdot 10^{-3}$ & $2.0\cdot 10^{-4}$ & $4.2\cdot 10^{-2}$ & $2.9\cdot 10^{-2}$\\
$100$ & $1.3$ & $2.2\cdot 10^{-2}$ & $2.9\cdot 10^{-2}$ & $9.6\cdot 10^{-4}$ & $4.5\cdot 10^{-2}$ & $3.3\cdot 10^{-2}$\\
\end{tabular}
\end{ruledtabular}
\caption{\label{tab:numbers2}
Characteristic numbers for simulations with constant acceleration and reactivity enabled. The reaction-rate constant is $k=(1/4)H/N_{\mathrm{t}}$.
The results thus fulfill the condition $r(t=200000)\approx H/4$. $H$ is the initial width of the channel. $Q_{1}$ determines the relative strength
between gravitational and viscous forces. $Q_{2}$, $Q_{3}$ and $Q_{4}$ compare the reaction with gravitational, viscous and capillary forces.
$Bo$ is the Bond number comparing gravity and surface tension; $Ca$ is the capillary number comparing surface tension and viscosity. For 
a given initial height an average is taken over the different applied accelerations, chosen as in Fig.~\ref{fig:reaction}. The time-varying
quantities are averaged inside a window centered around the time $N_{\mathrm{t}}/2=100000$ [ts].}
\end{table}
\begin{figure}[t]
\includegraphics[width=8.4cm]{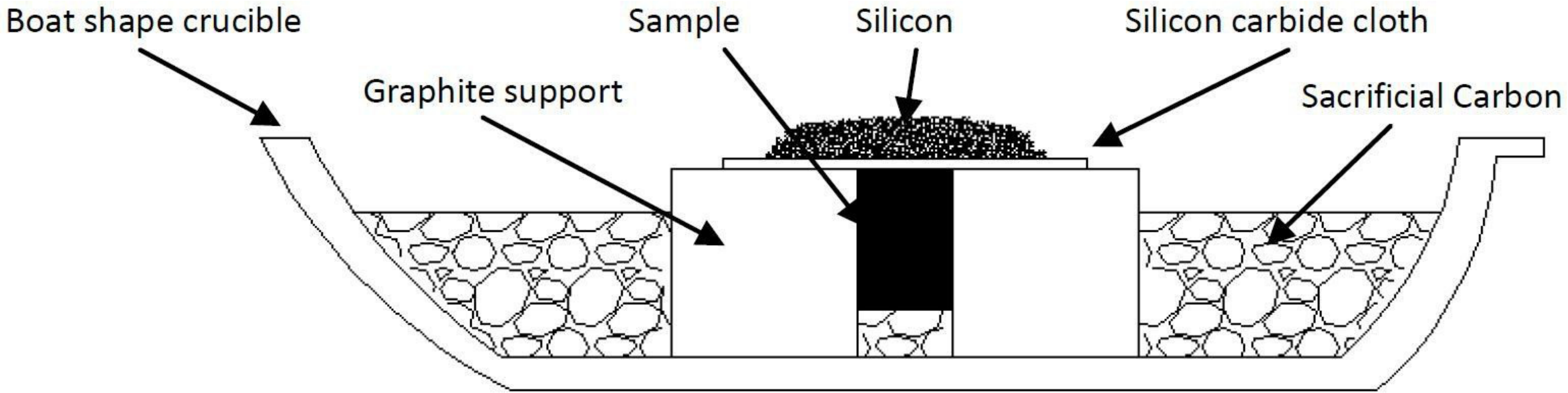}\\
\includegraphics[width=3cm]{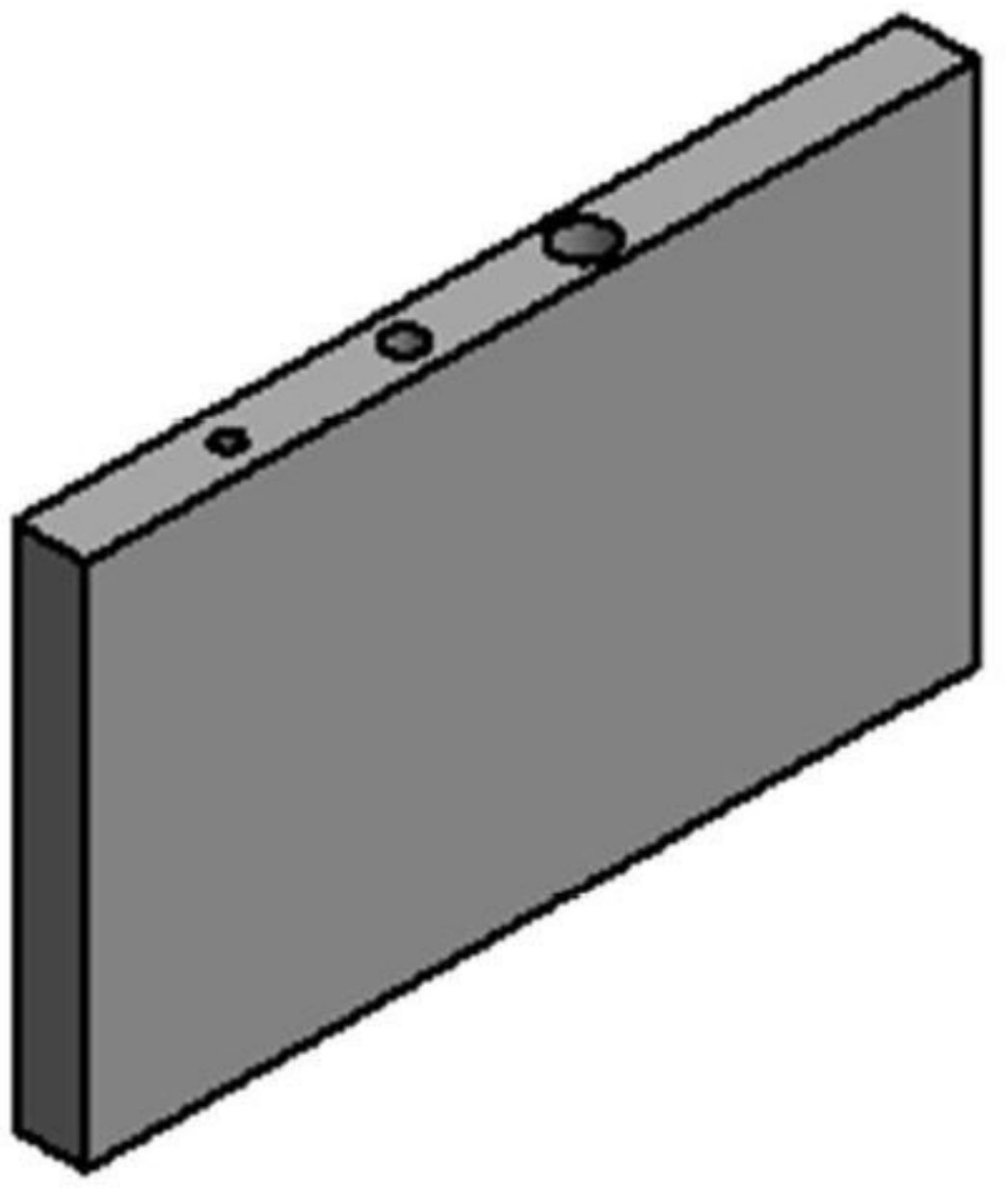}
\caption{\label{fig:material}
Top: Designed setup for experiments. The description and function of the indicated parts can be found in main text
(see Sec.~\ref{sec:material}). Bottom: Schematic for the sample with drilled channels.}
\end{figure}
\begin{figure}[t]
\includegraphics[width=8.4cm]{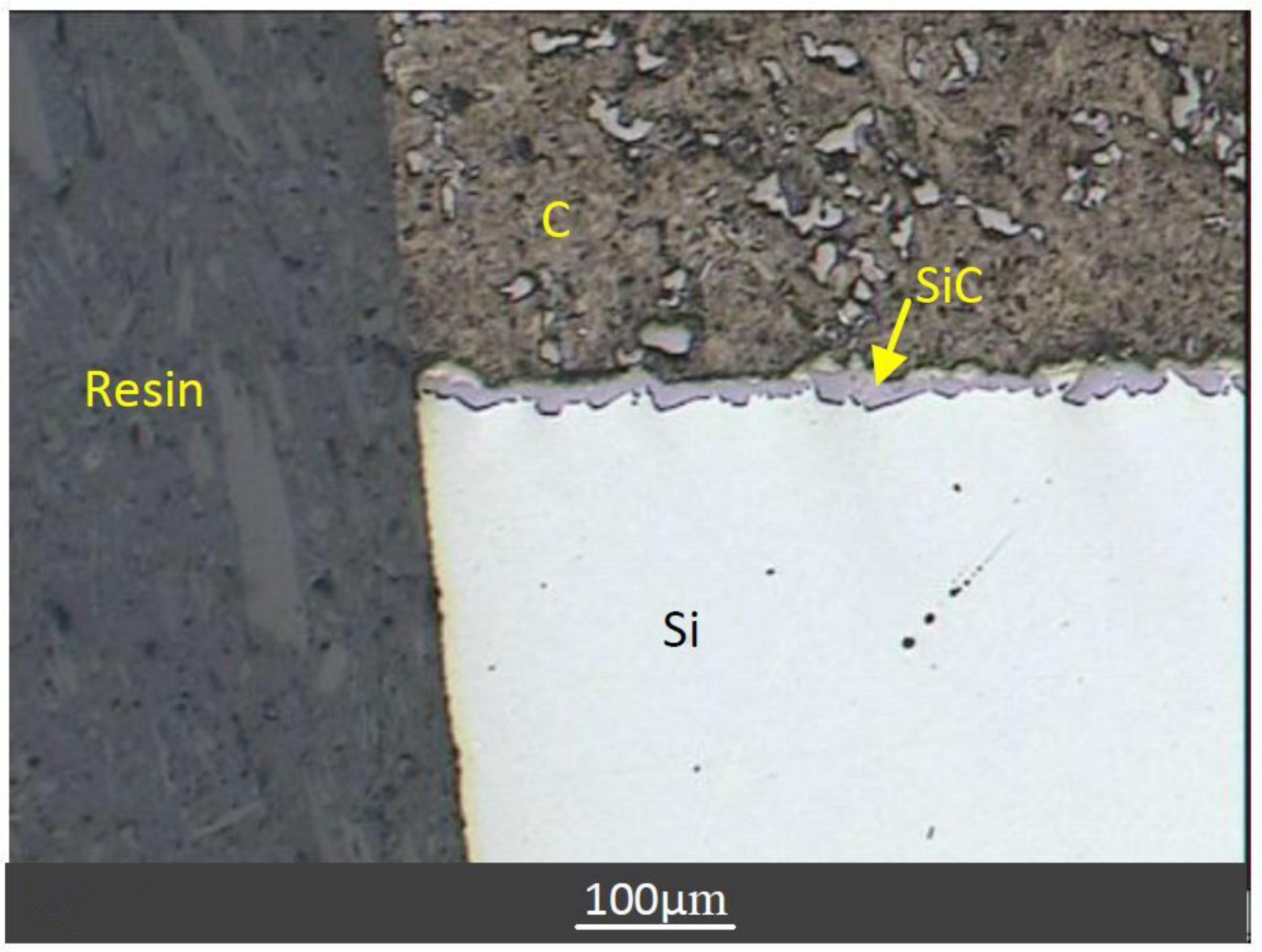}
\caption{\label{fig:exp1}
Micrograph near the inlet for the channel of diameter 1 [mm] infiltrated with a dwell time of 5 [min].}
\end{figure}
\begin{figure}[t]
\includegraphics[width=8.4cm]{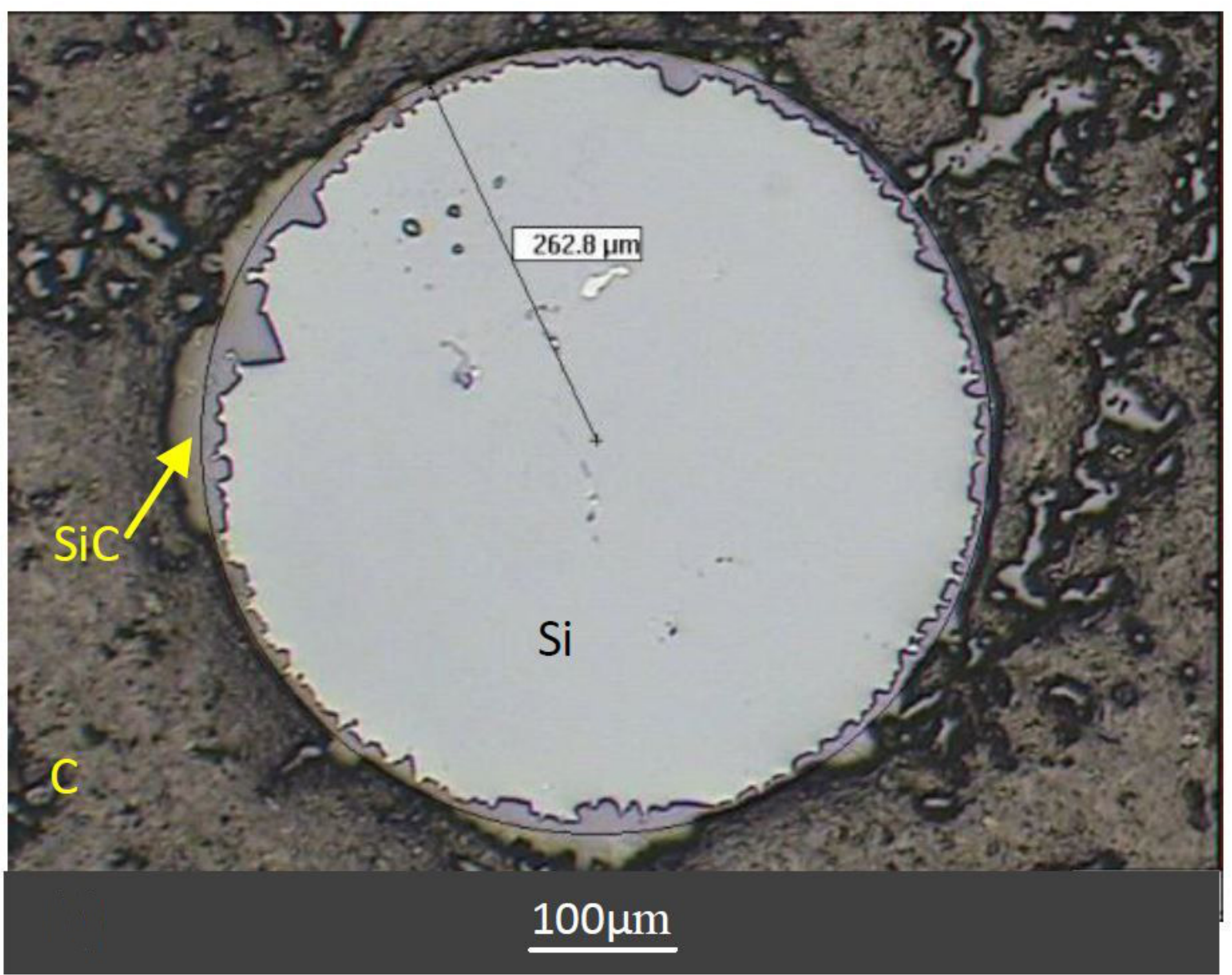}
\caption{\label{fig:exp5}
Cross-section of infiltrated channel of diameter $0.5$ [mm] with dwell time of $24$ hours. The circle represented with a solid line
indicates the initial diameter. The process of C dissolution leads to an increase of the diameter to $9.2\pm 4$ [$\mu$m] on average.}
\end{figure}
\begin{figure}[t]
\includegraphics[width=8.4cm]{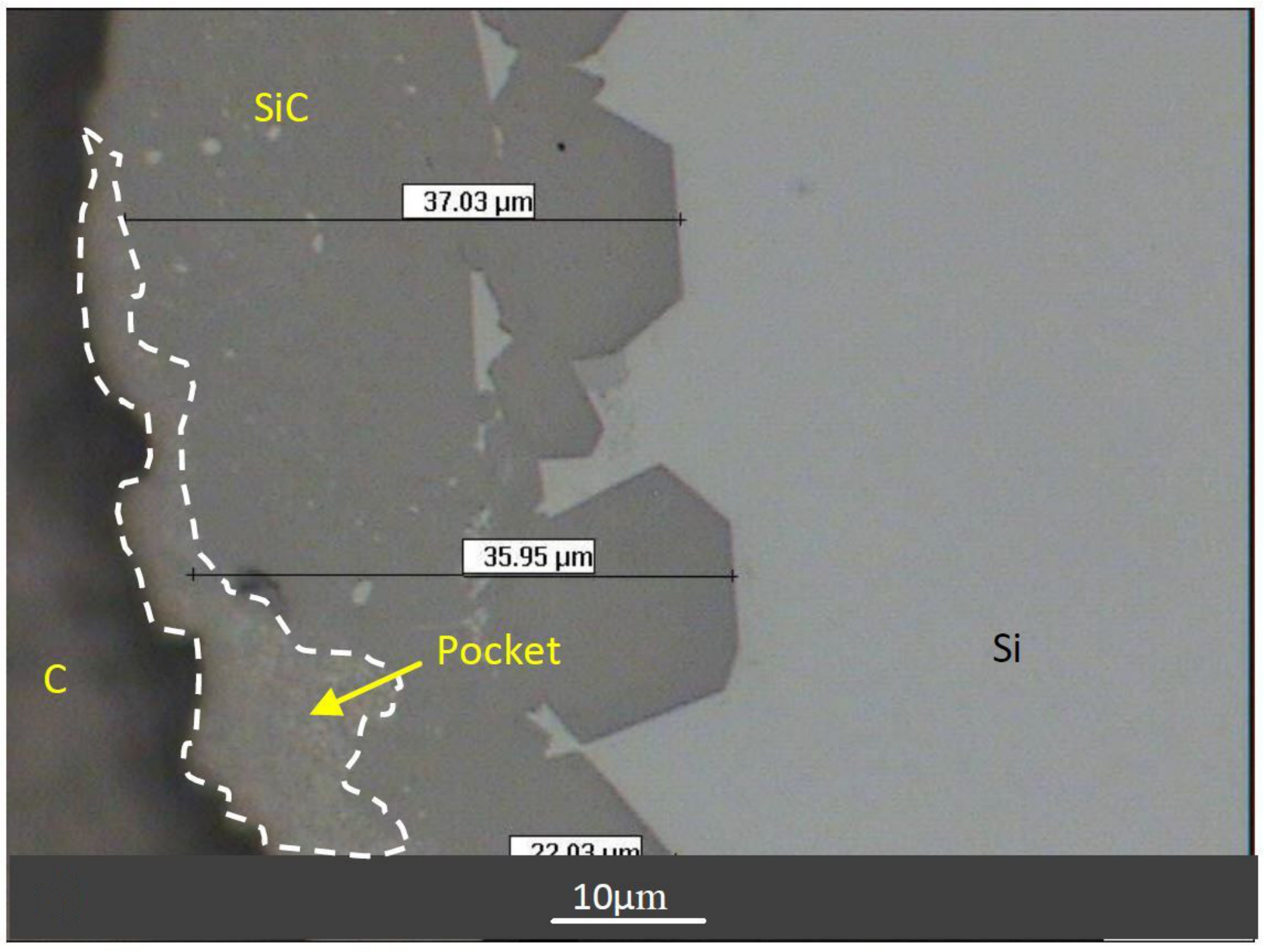}
\caption{\label{fig:pocket}
Formation of pockets in the growth of SiC. The initial diameter of the channel is $0.5$ [mm] and the experiment proceeded
with a dwell time of $24$ hours.}
\end{figure}
\begin{figure}[t]
\includegraphics[width=8.4cm]{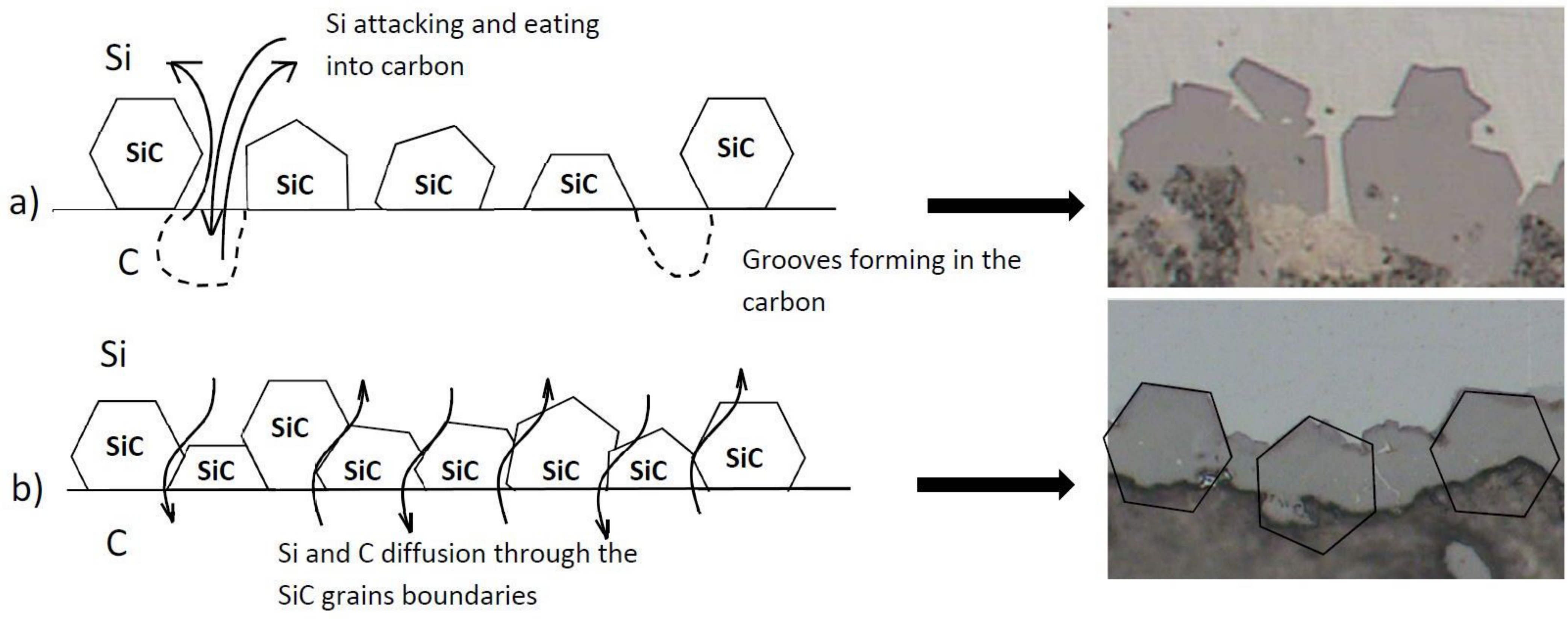}
\caption{\label{fig:process}
Schematic representation of the two separate stages at the origin of SiC growth with the formation of pockets \cite{pocket}.}
\end{figure}
\begin{figure}[t]
\includegraphics[width=8.4cm]{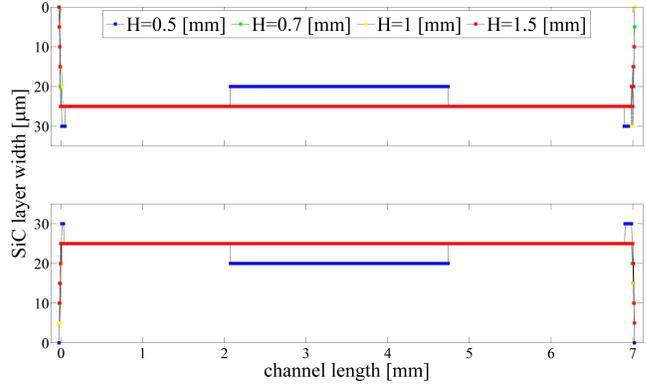}
\caption{\label{fig:test}
Profile of the SiC layer as resulting from simulations after $12.5$ [min]. As the height of the channels triples, the
width of the reaction-formed phase can be assumed to remain comparable, as in experiments. Top: upper wall; Bottom: lower
wall. $H$ is the initial height of the channels, i.e.~the initial diameter. The initial length of the channels is
$L=7$ [mm].}
\end{figure}

\subsection{Infiltration driven by weight with surface reaction}

We now enable the reaction leading to surface growth. The initial conditions for solute transport are $C_{1}=1.2\cdot 10^{-2}$ 
[mu/lu$^{2}$] in the liquid phase and $C_{2}=7.5\cdot 10^{-2}$ [mu/lu$^{2}$] in the vapor phase. The saturated concentration
is set to $C_{\mathrm{s}}=5\cdot10^{-3}$ [mu/lu$^{2}$]. The parameters for the function of Eq.~\ref{eq:fs} introducing
an interface are given by $G_{\mathrm{s}}=-4.875\cdot 10^{-3}$ [mu/lu/ts$^{2}$], $\varphi_{0}=1$ 
and $C_{0}=4.9\cdot 10^{-3}$ [mu/lu$^{2}$] \cite{supsi1,supsi2}. The initial mass on the solid
boundaries is $b_{0}=2\cdot 10^{-3}$ [mu]. To start with, the reaction-rate constant is set to $k=(1/2)H/N_{\mathrm{t}}$, where
$N_{\mathrm{t}}=200000$ is the total number of iterations. This condition alone is not sufficient in order to ensure that
pore clogging occurs at time $N_{\mathrm{t}}$, since there remains to be fixed the threshold value $b_{\mathrm{max}}$ for the 
cumulated mass. Figure \ref{fig:reaction} shows the results for the suitable values of $b_{\mathrm{max}}$. 
For two systems the surface grows more slowly than expected. This turns out to be related to the formation of the interface
for solute near the opening of the channel preventing solute from reaching properly the solid surface. This shortcoming can
be reduced by considering higher initial values for the solute concentration in the liquid phase. For comparative purposes we
keep the present settings. In Tab.~\ref{tab:numbers} are listed the results for the characteristic numbers. 
The values for $Q_{2}$, $Q_{3}$ and $Q_{4}$ are representative for systems dominated by reactivity, as it is evident from
Fig.~\ref{fig:reaction}.
The reactivity develops a stronger influence as the width of the channel doubles with an increase of $Q_{2}$, $Q_{3}$ and $Q_{4}$
approximately by a factor of $4$.
Gravity remains the dominant force: an estimate of the terms due to gravitational and capillary forces in the differential
equation reveals that the latter is $9.6$ times smaller (see Eq.~\ref{eq:eqdiff}).
Comparison with the results without reaction, Tab.~\ref{tab:numbers0}, indicates that the viscosity
becomes stronger with respect to gravity. Furthermore, surface tension becomes more relevant with respect to 
gravity. We thus conclude that the theoretical predictions shown in Fig.~\ref{fig:reaction}, including also capillary effects, 
are closer to the simulation results for this reason. Actually, we could estimate that for $H=50$ [lu], in the absence of reactivity, 
the term due to surface tension is $15$ times smaller than that introducing a constant acceleration.
It is also found that the spatial variations of the density for the simulations of Fig.~\ref{fig:reaction} are comparable to those
of Fig.~\ref{fig:g}, where the accordance with the theory is reasonable. This means that the treatment of the forcing term does not
represent a significant source of inaccuracy \cite{guo}.
In Fig.~\ref{fig:reaction}, the infiltration resulting from simulations is still faster presumably because it is assumed that the 
channels shrink uniformly. The fact that at the contact line the width of the channel is the initial one is in particular relevant 
in 3D (see Eq.~\ref{eq:capillary}). The results for $H=50$ [lu] clearly show that surface growth can not be disregarded
when the reduction of the radius is larger than $1/2$. Thus, we consider surface reaction as the dominant phenomenon. The best 
precision is obtained for $H=50$ [lu], when inertia forces are still weak at the beginning. 
In Tab.~\ref{tab:numbers} we report the characteristic numbers for the condition $k=(1/4)H/N_{\mathrm{t}}$. In words, this means
that at time $N_{\mathrm{t}}$, the end of the simulation, the initial radius is divided by $2$. It can be seen that gravity becomes
more relevant than in the previous case as indicated by $Q_{1}$, $Q_{2}$ and $Bo$. Also viscosity becomes more significant with respect
to the reaction. The relative strength among the other forces remains almost unchanged. Again, $Q_{2}$, $Q_{3}$ and $Q_{4}$ increase 
by a factor of $4$ as the initial height doubles. 
In Fig.~\ref{fig:equivalence} we verify the equivalence of systems based on the characteristic number $Q_{2}$. The agreement between 
the curves is good. This representation confirms that the reactivity and gravity account for the dominant effects. As discussed
before, the discrepancy with the theoretical results can be ascribed to the fact that capillary forces remains significant. It should
also be noted that in the representation of the results of Fig.~\ref{fig:reaction} the difference turns out to be of $3\%$.


\section{Experiments}

\subsection{Material, setup and method}\label{sec:material}

Figure \ref{fig:material} shows the setup egineered for the experiments of reactive infiltration into millimeter-sized channels.
The graphite used in order to prepare the preforms was provided by Schunk Graphite Technology. The porosity of the samples is
constrained below $6\%$, the density is $1.94$ [g/cm$^{2}$] with very low ash content. As-received Si is a powder consisting of flakes
at most of $500$ [$\mu$m] with a purity of $99.9\%$. Fine control of the infiltration process is achieved by making use of alumina
boat-shaped crucibles for the assembly of the components (see Fig.~\ref{fig:material}).
High-density alumina is a common choice for high-temperature technologies due to its
thermal, mechanical and chemical resistance. A SiC cloth is prepared as support for Si flakes and in order to filtrate Si oxides
at the time of infiltration. The auxiliary elements are in graphite coated with boron nitride so as to prevent molten Si from reacting.
Undesired effects of residual oxygen are avoided by surrounding the sample with activated carbon particulate with size less than $1$ [mm].
The experiments are performed in a horizontal tube furnace at a temperature of $1450^{\circ}$ [C] (heating rate of $3^{\circ}$ [C/min]) under
argon atmosphere (flow rate of $100$ [cm$^{3}$/min]). The infiltrated samples are sectioned and polished for surface analysis
with optical microscopy. The channels are realized in the compact graphite with a drill press or with a milling machine for
the smallest diameter of $0.5$ [mm]. Besides compressed air, the cleaning treatment of the channels with smallest diameter
consists also of ethanol ultrasonic bath. All the channels have a length of $7$ [mm]. The diameter of the channels and the dwell
time vary in the experiments.


\subsection{Experimental results}

The first set of experiments include channels of diameters of $0.7$, $1$ and $1.5$ [mm] for hold times of
$1$ hour, $30$ [min] and $5$ [min]. It turns out that the microstructure of all samples presents the same
morphology and the same rate of growth for SiC. It is observed that the width of the reaction-formed SiC
phase varies between $6$ and $25$ [$\mu$m] (cf.~Fig.~\ref{fig:exp1}). Inspection of the results indicates
that the thickness of the SiC layer is independent of the capillary diameter.

In the second series of experiments the diameter of the channels is fixed to $0.5$ [mm] and the dwell time is
increased to $24$ hours. The motivation for such a dwell time is to verify the slow process of SiC formation
following an initial growth. The obtained channels have a good circularity with the average diameter given by
$516\pm 3$ [$\mu$m]. After infiltration the diameter increases to an average value of $525.5\pm 1.5$ [$\mu$m],
as it can be seen in Fig.~\ref{fig:exp5}. This change in diameter of the channels can be associated with
the C dissolution process during the initial stage \cite{nisi2}. Also in this case, the crystals of SiC attain a thickness
between $6$ and $21$ [$\mu$m]. The only difference that can be identified is that the formed SiC is more compact
for longer dwell times. It is to be noted that SiC grows upward leaving behind its iterface regions filled with
Si, usually referred to as pockets. It turns out that during cooling, SiC crystals of submicrometric size
precipitate. Owing to C dissolution, the thickness of the SiC layer can
reach $38$ [$\mu$m], as shown in Fig.~\ref{fig:pocket}.

In another set of experiments the diameter of the channels is again $0.5$ [mm] and the dwell time is set to $72$ hours.
We point out that the SiC layer is not uniform along the wall of the channels. It is found that near the inlet the
thickness is between $20$ and $30$ [$\mu$m]. By advancing into the channels, the SiC thickness decreases up to reaching
a constant value in the range of $10-15$ [$\mu$m]. As before, the diameter of the capillaries after infiltration increases
to an average value of $524\pm 1$ [$\mu$m]. In general, the growth of the SiC layer is similar to that obtained with a dwell
time of $24$ hours. The main significant difference is that the SiC phase becomes more compact. Another differing aspect is that
the pockets become lower in number and more compact. This observation corroborates the mechanism of SiC formation
proposed in Ref.~\cite{pocket}. Namely, SiC growth proceeds in two steps. At the beginning, the growth is faster because of
the nucleation of SiC crystals close to the C surface (distance in the range of $10-20$ [$\mu$m]). Between these
nuclei, Si dissolves C which is transported through the liquid to the growing crystals. This explains the quite
regular distance between pockets formed at SiC-C interface. This process is shown in Fig.~\ref{fig:process}.
The second phase starts when the crystals merge. At this point the process is governed by the mechanism of
Si and C diffusion through SiC, slower than in the case of transport through the liquid after C dissolution by Si.
The result is that the overall reaction rate slows down drastically. This is why the difference in thickness of SiC is slight
for experiments performed during $24$ and $72$ hours.


\subsection{Comparison with LB simulations}

Equation \ref{eq:z_w} points out that in experiments the infiltration driven by gravity is a transient process.
Also in the experimental work, the column of fluid above the opening of the channels can be assumed to be comparable to
the length of the channels. As a consequence, the surface grows mainly under a regime of diffusion for mass transport.
This means that an equivalence with the simulations can be established via the Damkohler number $Da$ \cite{crystal}. In order
to estimate the experimental $Da$ numbers we consider for the characteristic length of the systems the side of an
equivalent square. Given the diffusion coefficient $D=5\cdot10^{-9}$ [m$^{2}$/s] \cite{nisi2} and the reaction-rate constant
$k=4\cdot10^{-8}$ [m/s] of Si, it is found that the $Da$ number varies in the range $0.015-0.026$. It follows
that the process of Si diffusion is much faster than the reactivity. In the LB systems, the length of the
channels is set to $L=1400$ [lu], corresponding to $7$ [mm] (see Sec.~\ref{sec:material}). Instead, the width of
the simulation domain is determined by the aspect ratio of the experimental channels, leading to $N_{y}=100,140,200,300$
[lu]. By reproducing the experimental $Da$ numbers, it is obtained that the reaction-rate constant for the simulations
is the same for every system and equal to $k=10^{-5}$ [lu/ts]. In the LB models for the reaction, the surface grows following
a linear relation with time. For this reason we can only simulate the process up to the formation of a continuous SiC
layer (see Fig.~\ref{fig:process}). The number of timesteps is derived by imposing that the surface grows for a distance
corresponding to $30$ [$\mu$m] during this time. It is found that the iterations amount to $6\cdot10^{5}$ [ts]. 
The simulation domain is $3000$ [lu] long. The channels start at $x=1000$ [lu] ($L=1400$ [lu]). The system is filled with
liquid and solute up to $x=2500$ [lu]. The channel is thus immersed in the liquid and the system remains at rest. This
state could have been reached after a previous phase of infiltration driven by an external acceleration. In experiments
this regime is very short (see Eqs.~\ref{eq:z1} and \ref{eq:z_w}). The motivation of our choice is to maintain the same
conditions for mass deposition. Indeed, tests show that a contact angle of $90^{\circ}$ results in the possible appearance
of the interface for solute near the walls.

Figure \ref{fig:test} shows the final profile of all channels. It is interesting to note that, as the height triples,
the results remain comparable. Furthermore, the target value of $30$ [$\mu$m] for the maximal width of the SiC layer
is always neared. These properties are in agreement with experiments. For the larger channels, the process of surface
growth is slower. A possible reason is that the walls are more distant from the centerline, where the concentration of
solute is higher. On the other hand, in this case, the SiC layer tends to be uniform because there is more solute available
in the systems. The simulations fail to reproduce a profile decreasing inside the channels. It is found that the surface
grows from the extremities to the center. As a consequence, a factor responsible for this shortcoming is that, in
simulations, there is an additional source of solute close to the right opening of the channels. It turns out that this
phenomenon is less marked with the smaller channel. In systems for porous structures the channels are quite narrower and
not blind. These undesired effects should be weaker also by the presence of advection induced by the advancing front.

Additional experiments are carried out in order to single out the causes leading to a discrepancy for the SiC profile.
The channels have a diameter of $1$ [mm] with openings of inlet and outlet, as in simulations. Furthermore, the preforms
are in glassy carbon since this carbon is characterized by a reduced roughness, porosity and reactivity (Alfa Aesar supplier).
The experimental conditions are the same as before, except for the temperature set to $1475^{\circ}$ [C] and the size of the Si
flakes (below $5$ [mm]). In the first experiment, the length of the channel is $7.85$ [mm] and the holding time is $2$ hours.
At the outlet, the flow is stopped by the formation of SiC crystals. It is found that the SiC growth around the capillary wall is
uniform, forming a layer $25-35$ [$\mu$m] thick. It turns out that the microstructure of the reaction-formed phase exhibits a
higher roughness. In the second experiment, the length of the channel is $5.71$ [mm] and the holding time is $4$ hours. In this
case, at the outlet the infiltration stops with the formation of a pendant droplet. It is observed that the SiC layer formed
of is more compact but there is no significant difference in thickness. Again, both phenomena can be ascribed to the slow
process of C diffusion through the SiC layer, affecting only the density. These results for channels with outlet compare better
with the simulations of Fig.~\ref{fig:test}. They also prove that the initial fast infiltration under the action of gravity has
a marginal role for surface growth, as assumed in simulations. 


\section{Resistance from chemical reaction in capillary infiltration}

Let us indicate by $F$ any force acting on the meniscus, i.e.~the interface, in Eq.~\ref{eq:washburn}. Let us 
assume that the radius shrinks as $r(t)=r_{0}-kt$. From the experiments we also know that $z=vt$ \cite{nisi2,htc09}. If
we replace into Eq.~\ref{eq:washburn} with inertia neglected, it is found that the resistance from the reaction has the
explicit form \cite{nisi2}
\begin{equation}
F=-2\pi r_{0}\gamma\cos\theta+8\mu\pi\Big(v^{2}+\frac{k\gamma \cos\theta}{4\mu}\Big)t\ .
\label{eq:F}
\end{equation}
Here $v$ is the interstitial velocity for a single channel.

Fluid flow through a porous medium is generally treated using the empirical result known as Darcy law.
It states that the following relation holds
\begin{equation}
u=-\frac{K}{\mu}\frac{\Delta P}{z}\ .
\label{eq:darcy}
\end{equation}
$K$ is the permealibity and $P$ the pressure. Here $u$ is the superficial velocity; by dividing it by the porosity,
the interstitial velocity is obtained (continuity of fluid flow). In the Hagen-Poiseuille model, the porous medium is
regarded as an assembly of parallel capillary tubes \cite{bear}. In this model, for reactive flow, the permeability
takes the form \cite{bear,messner}
\begin{equation*}
K=\frac{\varphi_{0}}{8r_{0}^{2}}(r_{0}-kt)^{4}=\frac{\varphi(t)}{8}r(t)^{2}\ .
\end{equation*}
$\varphi_{0}$ is the initial porosity. The porosity in the course of time is given by
$\varphi(t)=\varphi_{0}(r_{0}-kt)^{2}/r_{0}^{2}$. In order to integrate Eq.~\ref{eq:darcy} we also use the
relation
\begin{equation*}
z=\frac{3kt}{\varphi_{0}r_{0}[1-(1-\frac{k}{r_{0}}t)^{3}]}q\ .
\end{equation*}
$q$ is the distance allowing to calculate the fluid flow in the reservoir.
When pore closure occurs $z=3q/\varphi_{0}$, where $\varphi_{0}/3$ is the average porosity.

If the resistance arising from the chemical reaction of Eq.~\ref{eq:F} is taken into account, in Eq.~\ref{eq:darcy}
the pressure becomes $\Delta P=-8\mu\pi v^{2}t/\pi r^{2}$. Darcy law, Eq.~\ref{eq:darcy}, leads to
\begin{equation}
q^{2}-q_{0}^{2}=\frac{\varphi_{0}^{2}r_{0}^{2}v^{2}}{9k^{2}}\Big[1-\Big(1-\frac{k}{r_{0}}t\Big)^{3}\Big]^{2}\ .
\label{eq:darcy_q}
\end{equation}
The velocity $v$ is a representative value for the velocity inside the capillary tubes. The dependence on the
surface tension, the contact angle and the viscosity is incorporated into the velocity $v$. The infiltrated distance
is determined by the following integral ($\dot{q}$ derivative with respect to time)
\begin{eqnarray*}
z-z_{0}&=&\int^{t}_{0}\frac{\dot{q}}{\varphi}\mathrm{d}t\\
&=&\frac{\varphi_{0}r_{0}v^{2}}{3k}\int^{t}_{0}\frac{ 1-(1-\frac{k}{r_{0}}t)^{3} }{   
\sqrt{ \frac{\varphi_{0}^{2}r_{0}^{2}v^{2}}{9k^{2}} [1-(1-\frac{k}{r_{0}}t)^{3}]^{2}+q_{0}^{2} }}
\mathrm{d}t\ .
\label{eq:zinf}
\end{eqnarray*}
If $q_{0}=0$ we obtain the expected result $z=vt$ and thus Eq.~\ref{eq:F} is correct.


\section{Conclusions}

In this study, the focus is on millimeter-sized channels in order to elucidate relevant aspects of Si infiltration into C preforms.
A systematic series of experiments is performed with the purpose to detail the process of surface growth. The experimental work clearly
shows that the formation of SiC from the reaction between Si and C proceeds through two stages. The first one is mainly responsible for
the thickening of the SiC layer. This stage is characterized by diffusion of reactants in the liquid. In the second stage, the diffusion
of reactants occurs through the reaction-formed SiC solid phase. This mechanism is much slower and has the effect to make the SiC layer
more uniform and compact.

The system of a single channel has also the advantage to ease the comparison with simulations in order to improve the description of
the surface reaction \cite{supsi1,supsi2}. Dimensionless numbers are introduced with the aim to assess the accuracy of simulations
involving surface growth. Our analysis points out that the characteristic numbers lead to a consistent approach. This is critical
for extensions to porous structures. Furthermore, this work indicates that the models for the reaction support a proper implementation
of the pore-closing phenomenon. This means that the important parameter of the reaction-rate constant can be regarded as an effective
rate of growth. Constraints arising from experiments can be imposed for the profile of the SiC layer. There appears that in
the diffusive regime the structure of the growing phase can be regarded as a crude approximation of the microstructure of SiC.
It is to be noted that infiltration velocities are low for real porous preforms \cite{einset1,patro}.

Finally, this work represents a substantial step toward the simulation of reactive infiltration into porous structures. Our investigation
suggests that a correspondence between infiltrated distances in simulations and experiments can be established. Furthermore, the comparison
can be made more stringent by means of characteristic numbers taking into account the surface reaction. The properties of the forming
SiC layer reproduced by the models are also a positive indication. The prospect is to address the involved problems of maximal penetration,
SiC formation and residual Si content for the optimization of ceramic materials.


\acknowledgments

D.S. thanks Prof.~Narciso for a stay at the University of Alicante, as well as 
the members of his group for helpful discussions. The research leading to these 
results has received funding from the European Union Seventh Framework Programme 
(FP7/2007-2013) under grant agreement n$^{\circ}$ 280464, project "High-frequency 
ELectro-Magnetic technologies for advanced processing of ceramic matrix composites 
and graphite expansion'' (HELM).


\end{document}